\documentclass[final,3p,times]{elsarticle}
\usepackage{amsmath,amsfonts}
\usepackage{algorithmic}
\usepackage{array}
\usepackage[caption=false,font=normalsize,labelfont=sf,textfont=sf]{subfig}
\usepackage{textcomp}
\usepackage{stfloats}
\usepackage{url}
\usepackage{verbatim}
\usepackage{graphicx}

\usepackage{xcolor}
\usepackage{graphicx}
\usepackage{amssymb}
\usepackage{enumitem}
\usepackage{booktabs}

\usepackage[ruled,vlined,linesnumbered]{algorithm2e} 

\SetCommentSty{mycommfont}

\definecolor{darkgreen}{RGB}{0,104,0}

\usepackage{amsthm}

\newtheorem{theorem}{Theorem}[section]

\theoremstyle{definition}
\newtheorem{definition}{Definition}[section]
\theoremstyle{definition}
\newtheorem{example}{Example}[section]
\usepackage{xcolor}

\newcommand{\degoutp} {{\it{d^+_{out}} }}

\newcommand{\degins} {{\it{d^*_{in}}}}
\newcommand{\od}{\mathbb{O}}  
\newcommand{\wP}{\mathcal{P}}
\newcommand{\vgr}{{\it{V^+ \setminus V^*}}}

\newcommand{\fs}{}
\newcommand{\adj}{{\it adj}}
\newcommand{\Deg}{{\it deg}} 
\newcommand{\dequeue}{{\it dequeue}}
\newcommand{\enqueue}{{\it enqueue}}

\newcommand{\updateversion}{{\it update\_version}}
\newcommand{\core}{{\it core}}

\newcommand{\post}{{\it{post}}}
\newcommand{\pre}{{\it{pre}}}
\newcommand{\mcd}{{\it{mcd}}}

\newcommand{\group}{{\it group}}
\newcommand{\lock}{{\it lock}}
\newcommand{\Next}{{\it{next}}}

\newcommand{\version}{{\it ver}}
\newcommand{\counter}{{\it cnt}}

\newcommand{\alg}[1]{\textsc{#1}} 

\usepackage{todonotes}

\setitemize[0]{}
\setitemize[1]{}

\newcommand{\myfig}{Fig.~}
\usepackage{enumitem}
\usepackage{blindtext}

\setlist[itemize]{
}

\newif\iftpds 
\tpdstrue 

\journal{Journal of Parallel and Distributed Computing}
\begin{document}
\begin{frontmatter}



\title{Parallel Order-Based Core Maintenance in Dynamic Graphs}




\author[label1]{Bin Guo\corref{cor1}}
\ead{binguo@trentu.ca}
\cortext[cor1]{Corresponding author}
\affiliation[label1]{organization={Department of  Computing \& Information Systems, Trent University},
            addressline={1600 West Bank Drive}, 
            city={Peterborough},
            postcode={K9L 0G2}, 
            state={ON},
            country={Canada}}

 \author[label2]{Emil Sekerinski}
 \ead{emil@mcmaster.ca}
\affiliation[label2]{organization={Department of Computing and Software, McMaster University},
            addressline={1280 Main Street West}, 
            city={Hamilton},
            postcode={L8S 4L8}, 
            state={ON},
            country={Canada}}           

\begin{abstract}
The core number of vertices in a graph is one of the most well-studied cohesive subgraph models due to linear running-time algorithms. In practice, many data graphs are dynamic graphs that continuously change by inserting and removing edges. The core numbers are updated in dynamic graphs with edge insertions and deletions, called core maintenance. When a burst of a large number of inserted or removed edges comes in, these must be handled on time to keep up with the data stream. There are two main sequential algorithms for core maintenance, \textsc{Traversal} and \textsc{Order}. Experiments show that the \textsc{Order} algorithm significantly outperforms the \textsc{Traversal} algorithm over various real graphs. 

To the best of our knowledge, all existing parallel approaches are based on the \textsc{Traversal} algorithm. These algorithms exploit parallelism only for vertices with different core numbers; they reduce to sequential algorithms when all vertices have the same core numbers. This paper proposes a new parallel core maintenance algorithm based on the \textsc{Order} algorithm. A distinguishing properly is that the algorithm allows parallelism even for graphs where all vertices have the same core number. Extensive experiments are conducted over real-world, temporal, and synthetic graphs on a multicore machine. The results show that for inserting and removing a batch of edges using 16 workers, the proposed method achieves speeds of up to 289 times and 10 times compared to the most efficient existing method.

\end{abstract}




\end{frontmatter}


\section{Introduction}


Graphs are widely used to model complex networks, such as social networks, hyperlink networks, and knowledge networks.   
As one of the most well-studied cohesive subgraph models, the $k$-core is the maximal subgraph such that all vertices have degrees at least $k$. 
The \emph{core number} of a vertex is the maximum value of $k$ such that this vertex is contained in the subgraph of $k$-core vertices~\cite{bz2003,Kong2019}.
The core numbers can be computed in linear time $O(m)$ by the \alg{BZ} algorithm \cite{bz2003}, where $m$ is the number of edges in a graph.
Due to such computational efficiency, the core number of a vertex is a parameter of density extensively used in numerous applications~\cite{Kong2019}, such as knowledge discovery~\cite{wang2022knowledge}, gene expression~\cite{dorantes2021k}, social networks~\cite{gao2020k}, ecology~\cite{burleson2020k}, and finance~\cite{burleson2020k}.

In~\cite{Malliaros2020}, Malliaros et al. summarize the main research on $k$-core decomposition from 1968 to 2019.
Many papers focus on computing the core in static graphs~\cite{bz2003,cheng2011efficient,khaouid2015k,montresor2012distributed,wen2016efficient}.
In practice, many data graphs are large and continuously changing. It is of practical importance to identify the dense range as fast as possible after a change, e.g., when multiple edges are inserted or removed.
For example, it is necessary to quickly initiate a response to rapidly spreading false information about vaccines or to urgently address new pandemic super-spreading events~\cite{miorandi2010k,pei2014searching,gabert2021shared}. 
This is the problem of maintaining the core number in dynamic graphs. 
In~\cite{zhang2020core}, Zhang et al.~summarize the research on core maintenance and applications. 

Many sequential algorithms are devised for core maintenance in dynamic graphs~\cite{guo2022simplified,Zhang2017,Saryuce2016,wu2015core,sariyuce2013streaming,li2013efficient}. 
The main idea for core maintenance is that first, a set of vertices whose core numbers need to be updated (denoted as $V^*$) is identified by traversing a possibly larger scope of vertices (denoted as $V^+$) such that $V^* \subseteq V^+$.
There are two main algorithms for maintaining core numbers over dynamic graphs, \alg{Traversal}~\cite{Saryuce2016} and \alg{Order}~\cite{Zhang2017}.
Given an inserted edge, the \alg{Traversal} algorithm searches $V^*$ by performing a depth-first graph traversal within a \emph{subcore}, a connected region of vertices with the same core numbers. 
For the \alg{Order} algorithm, the size of $V^+$ is significantly reduced, so the ratio $|V^+|/|V^*|$ is typically much smaller and has less variation compared to the \alg{Traversal} algorithm. Thus, the computational time is significantly improved.
The experiments in~\cite{Zhang2017} show that for edge insertion, \alg{Order} significantly outperforms \alg{Traversal} over all tested graphs with up to 2,083 times speedups; for the edge removal, \alg{Order} outperforms \alg{Traversal} over most of the tested graphs with up to 11 times speedups. 
Furthermore, based on the \alg{Order} algorithm, a \alg{Simplified-Order} algorithm is proposed in~\cite{guo2022simplified}, which is easier to implement and to argue for correctness; also, \alg{Simplified-Order} has improved time complexities by adopting the \emph{Order Maintenance} (OM) data structure to maintain the order of all vertices.

All the above methods are sequential for maintaining core over dynamic graphs, meaning they handle only one edge insertion or removal at a time. 
The problem is that when a burst of many inserted or removed edges comes in, these edges may not be handled on time to keep up with the data stream~\cite{gabert2021shared}.
The prevalence of multi-core machines suggests parallelizing the core maintenance algorithms.
Many multi-core parallel batch algorithms for core maintenance have been proposed in~\cite{hua2019faster,Jin2018,wang2017parallel}. 
These algorithms are based on similar ideas: 1) they use an available structure, e.g.~\emph{Join Edge Set}~\cite{hua2019faster}, to preprocess a batch of inserted or removed edges, avoiding repeated computations, and 2) each worker performs the \alg{Traversal} algorithm.
However, there are three drawbacks to these approaches. First, they are based on the sequential \alg{Traversal} algorithm~\cite{sariyuce2013streaming,li2013efficient}, which is proved to be less efficient than the \alg{Order} algorithm~\cite{hua2019faster,guo2022simplified}. Second, parallelism can be exploited only for affected vertices with different core numbers; the algorithm is reduced to running sequentially when all affected vertices have the same core numbers. Third, the time complexities are not analyzed by the \emph{work-depth} model \cite{cormen2022introduction}, where the work $\mathcal W$ is its sequential running time, and the depth $\mathcal D$ is its running time on an infinite number of processors. 


    
    

To overcome the above drawbacks, inspired by the \alg{Simplified-Order} algorithm~\cite{guo2022simplified}, we propose a new parallel batch algorithm, called \alg{Parallel-Order}, to maintain core numbers after batches of edges insertions or removals in dynamic graphs. 
Based on maintaining the $k$-order with the parallel OM data structure, we can parallelize the \alg{Order} algorithm. 1) For edge insertion, the idea is straightforward: each worker handles one inserted edge at a time and propagates the affected vertices in $k$-order, such that only the traversed vertices in $V^+$ are locked. 
When traversing $v\in V^+$, not all neighbors $u\in v.\adj$ need to be locked if $u \notin V^+$. 
When another worker tries to access these vertices, it must wait until they are unlocked. Since all affected vertices are locked in $k$ order, this method does not have blocking cycles that lead to a deadlock.
2) Edge removal is more challenging than edge insertion. The idea is that each worker handles one removed edge at a time and propagates the affected vertices, such that only the traversed vertices in $V^*$ are locked. 
When traversing $v\in V^*$, not all neighbors $u\in v.\adj$ need to be locked if $u \notin V^*$.
The problem is that this method may cause blocking cycles that lead to a deadlock as all affected vertices are traversed without any order. We propose a novel mechanism to avoid such deadlocks. 
The main contributions of this work are summarized below:
\begin{itemize} 
    \item We investigate the drawbacks of the state-of-the-art parallel core maintenance algorithms~\cite{hua2019faster,Jin2018,wang2017parallel}. 
    
    \item Based on the parallel OM data structure\cite{guo2022new} and the \alg{Simplified-Order} core maintenance algorithm\cite{guo2022simplified}, we propose a \alg{Parallel-Order} edge insertion algorithm for core maintenance. Only the traversed vertices in $V^+$ are locked for synchronization. 
    We also implement the priority queue $Q$ combined with the parallel OM data structure to obtain a vertex in $Q$ with a minimal $k$-order.

    \item We propose a \alg{Parallel-Order} edge removal algorithm for core maintenance and a novel mechanism to avoid blocking cycles that lead to a deadlock. Only the traversed vertices in $V^*$ are locked for synchronization. 
    
    \item We prove the correctness of our \alg{Parallel-Order} insertion and removal core maintenance algorithms; we also prove the time complexities with the work-depth model.
    
    \item We conduct extensive experiments on a 64-core machine over various graphs to evaluate the \alg{Parallel-Order} algorithms for edge insertion and removal. 
\end{itemize}

\iftpds
\begin{table*}[tbh]
\begin{center}

\caption{The work and depth complexities of our parallel core maintenance}
\begin{tabular}{l|cc | cc}
\hline
       & \multicolumn{2}{ c}{Worst-case ($O$) } &  \multicolumn{2}{|c}{Best-case ($O$)}   \\ 
  Parallel& $\mathcal W$ & $\mathcal D$ &  $\mathcal W$ & $\mathcal D$ \\ \hline 
  Insert    & $m' |E^+| \log |E^+|$ & $m' |E^+| \log |E^+|$    & $m' |E^+| \log |E^+|$  & $|E^+| \log |E^+| + m'|V^*|$  \\
  Remove    & $m' |E^*|$           &$m' |E^*|$      &    $m' |E^*|$      &     $|E^*|+m'|V^*|$         \\
 \hline
\end{tabular}
\end{center}

\label{tab:comp}
\end{table*} 
\fi 

\begin{figure}[htb]
\centering
\includegraphics[width=\linewidth]{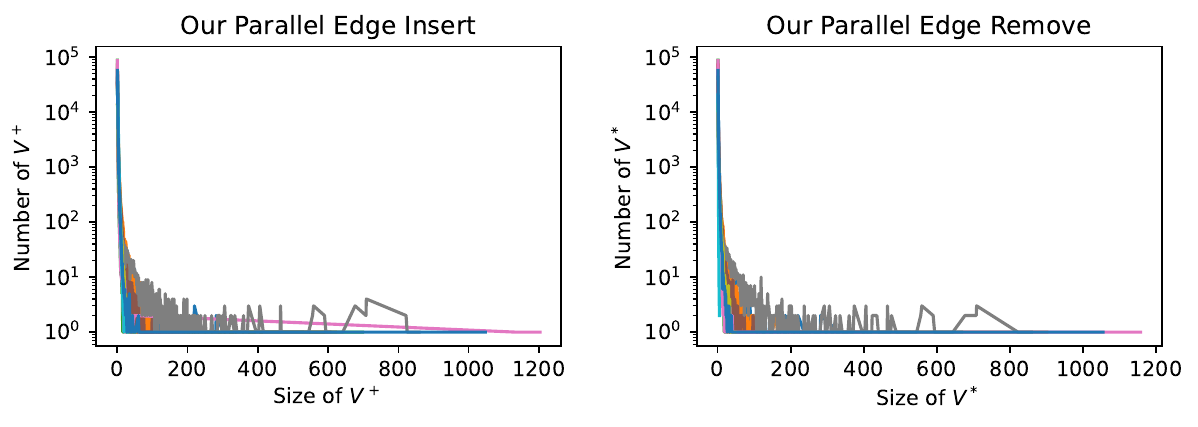} 
\centering
\caption{\rm The number of different sizes of $V^+$ for inserting and removing 100,000 edges by using our parallel core maintenance algorithms, respectively.
The x-axis is the size of $V^+$ and the y-axis is the number of such size of $V^*$. }
\label{fig:vcolor}
\end{figure}

We analyze our parallel algorithms in the standard \emph{work-depth} model~\cite{jeje1992introduction}. 
The \emph{work}, denoted as $\mathcal W$, is the total number of operations that the algorithm uses.
The \emph{depth}, denoted as~$\mathcal D$, is the longest chain of sequential operations. 
\iftpds
Table \ref{tab:comp} shows the work and depth of the \alg{Parallel-Order} algorithm for inserting and removing $m'$ edges in parallel. 
\fi
For both edge insertion and removal, one issue is that the depth $\mathcal D$ is equal to the work $\mathcal W$ in the worst case; that is, all workers execute in one blocking chain such that only one worker is active. 
However, with a high probability, such a worst-case does not happen as the number of locked vertices is always small for each insertion and removal. 
For our method, all vertices in $V^+$ or $V^*$ are locked together for each insertion or removal. 
In \myfig\ref{fig:vcolor}, on 16 tested graphs (in Table~\ref{tb:graph} of our experiment section), we summarize the number of different sizes of $V^+$ when randomly inserting and removing 100,000 edges. 
We observe that almost all $V^+$ and $V^*$ have really small sizes for insertion or removal, respectively. 
Specifically, more than 97\% of edge insertions and removals for all tested graphs have $|V^+|$ and $|V^*|$ between 0 and 10. That means, with a high probability that less than 11 vertices are locked when inserting or removing one edge, which leads to a low probability that all workers execute in one long blocking chain.

The rest of this paper is organized as follows. The related work is discussed in Section~\ref{relatedwork}. The preliminaries are given in Section~\ref{Preliminaries}. 
Our new parallelized Order-Based core maintenance algorithms are proposed in Section~\ref{parallel-core-maint}. 
We discuss the implementation of priority queues and buffer queues in Section~\ref{implementation}.   
We conduct extensive performance studies and show the results in Section~\ref{experiments}, and conclude this work in Section~\ref{conclusions}.

\section{Related Work}
\label{relatedwork}
\subsection{Core Decomposition}
\noindent
The BZ algorithm \cite{bz2003} has linear running time $O(m)$ by using bucket structures, where $m$ is the number of edges. 
In \cite{cheng2011efficient}, an external memory algorithm is proposed, so-called EMcore, which runs in a top-down manner such that the whole graph does not have to be loaded into memory. In~\cite{wen2016efficient}, Wen et al.~provide a semi-external algorithm, which requires $O(n)$ memory to maintain the information of vertices, where $n$ is the number of vertices. In \cite{khaouid2015k}, Khaouid et al. investigate the core decomposition in a single PC over large graphs by using  \texttt{GraphChi} and \texttt{WebGraph} models. In \cite{montresor2012distributed}, Montresoret et al.~consider the core decomposition in a distributed system. 
In addition, the parallel computation of core decomposition in multi-core processors is first investigated in \cite{dasari2014park}, where the ParK algorithm was proposed. Based on the main idea of ParK, a more scalable PKC algorithm has been reported in \cite{kabir2017parallel}.

\subsection{Core Maintenance}

\noindent
In~\cite{sariyuce2013streaming,li2013efficient}, an algorithm similar to the \alg{Traversal} algorithm is given, but with a quadratic time complexity.
In~\cite{wen2016efficient}, a semi-external algorithm for core maintenance is proposed to reduce the I/O cost, but this method takes more CPU time than the \alg{Traversal} algorithm.
In~\cite{sun2020fully}, Sun et al.~design algorithms to maintain approximate cores in dynamic \emph{hypergraphs} in which a \emph{hyperedge} may contain one or more participating vertices compared with exactly two in graphs.
In~\cite{gabert2021shared}, Gabert et al. propose parallel core maintenance algorithms for maintaining cores over {hypergraphs}. 
There exists some research based on the above core maintenance. In~\cite{yu2021querying}, the authors study computing all $k$-cores in the graph snapshot over the time window. In~\cite{lin2021hierarchical}, the authors explore the hierarchy core maintenance. 
In~\cite{weng2021efficient}, distributed core maintenance is explored. 
In~\cite{liu2021parallel2}, a parallel approximate $k$-core decomposition and maintenance approach is proposed, where bounded approximate core numbers for vertices can be maintained with high probability.

\iftpds
\subsection{Weighted Graphs}
\noindent
All the above work focuses on unweighted graphs, but graphs are weighted in many applications.
For an edge-weighted graph, the degree of a vertex is the sum of the weights of all its incident edges.
However, weighted graphs have a large search range for maintaining the core numbers after a change by using the traditional core maintenance algorithms directly, as the degree of a related vertex may change widely. 
In~\cite{zhou2021core}, Zhou et al. extend the coreness to weighted graphs and devise weighted core decomposition algorithms; also, they devise weighted core maintenance based on the $k$-order~\cite{Zhang2017,guo2022simplified}. 
In~\cite{liu2020Increamental}, Liu et al. improve the core decomposition and incremental maintenance algorithm to suit edge-weighted graphs.
\fi 

\section{Preliminaries} \label{Preliminaries}
\noindent
Let $G = (V, E)$ be an undirected and unweighted graph; $V(G)$ denotes the set of vertices, and $E(G)$ represents the set of edges in $G$. When the context is clear, we will use $V$ and $E$ instead of $V(G)$ and $E(G)$, respectively.
As $G$ is an undirected graph, an edge $(u, v)\in E(G)$ is equivalent to $(v, u)\in E(G)$. 
We denote the number of vertices and edges of $G$ by $n$ and $m$, respectively. We define the set of neighbors of a vertex $u \in V$ as $u.\adj$, formally $u.\adj = \{v \in V: (u, v) \in E\}$.
We denote the degree of $u$ in $G$ as $u.\Deg = |u.\adj|$.

\begin{definition} [$k$-Core]
Given an undirected graph $G=(V, E)$ and an integer $k$, a subgraph $G_k$ of $G$ is called a $k$-core if it satisfies the following conditions: (1) for $\forall u \in V(G_k)$, $u.\Deg \geq k$; (2) $G_k$ is maximal. Moreover, $G_{k+1} \subseteq G_k$, for all $k \geq 0$, and $G_0$ is just $G$. 
\end{definition}

\begin{definition}[Core Number]
\label{def:corenumber}
Given an undirected graph $G=(V,E)$, the core number of a vertex $u\in G(V)$, denoted as $u.\core$, is defined as $u.\core = max\{k: u \in V(G_k)\}$. That means $u.core$ is the largest $k$ such that there exists a $k$-core containing $u$.
\end{definition}

\begin{definition} [$k$-Subcore]
Given an undirected graph $G=(V,E)$, a maximal set of vertices $S\subseteq V$ is called a $k$-subcore if (1) $\forall u \in S, u.core = k$; (2) the induced subgraph $G(S)$ is connected. The subcore that contain vertex $u$ is denoted as \texttt{sc}(u).  
\end{definition}

\subsection{Core Decomposition}
\noindent
Given a graph $G$, the problem of computing the core number for each $u \in V(G)$ is called core decomposition. In~\cite{bz2003}, Batagelj et al. propose a linear time $O(m+n)$ algorithm, the so-called BZ algorithm, shown in Algorithm \ref{alg:bz}. 
The general idea is \emph{peeling}: to compute the $k$-core $G_k$ of $G$, repeatedly vertices (and their adjacent edges) whose degrees are less than $k$ are removed. When there are no more vertices to be removed, the resulting graphs are the $k$-core of $G$. 
The core number of $u$ is determined in line 5. 
The min-priority queue $Q$ can be efficiently implemented by bucket sorting \cite{bz2003}, leading to a linear running time of $O(m+n)$.

\iftpds
\iftrue
\begin{algorithm}[tb]
\caption{BZ algorithm for core decomposition}
\label{alg:bz}
\SetAlgoNoEnd
\DontPrintSemicolon
\SetKwInOut{Input}{input}\SetKwInOut{Output}{output}
    
    \lFor{$u \in V$}{ 
        $u.d \gets |u.\adj|$; $u.core = \varnothing$}
    $Q\gets$ a min-priority queue by $u.d$ for all $u\in V$\;
    
    \While{$Q \neq \emptyset$}{
        $u \gets Q.\dequeue ()$\;
       $u.\core \gets u.d$; remove $u$ from $G$\; \label{alg:bz-core}
       \For{$v \in u.\adj$}{
            \lIf{$u.d < v.d$}{$v.d \gets v.d - 1$}
        }
        
        update $Q$\;
    }
\end{algorithm}
\fi 
\fi 

\subsection{Core Maintenance}
\noindent
The problem of maintaining the core numbers for dynamic graphs $G$ when edges are inserted into and removed from $G$ continuously is called core maintenance. The insertion and removal of vertices can be simulated as a sequence of edge insertions and removals. 

\begin{definition}[Candidate Set $V^*$ and Searching Set $V^+$]
Given a graph $G=(V,E)$, when an edge is inserted or removed, a candidate set of vertices, denoted $V^*$, needs to be identified, and the core numbers of vertices in $V^*$ must be updated. 
To identify $V^*$, a minimal set of vertices, denoted $V^+$, is traversed by accessing their adjacent edges.
\end{definition}

Clearly, we have $V^* \subseteq V^+$. An efficient core maintenance algorithm should have a small ratio of $|V^+|/|V^*|$. It is shown that the \alg{Order}~\cite{hua2019faster} insertion algorithm has a significantly lower ratio compared to the \alg{Traversal}~\cite{sariyuce2013streaming} insertion algorithm. This is why we try to parallelize the \alg{Order} algorithm in this paper. 

In \cite{li2013efficient,sariyuce2013streaming}, it is proved that after inserting or removing one edge, the core number of vertices in $V^*$ increases or decreases by at most one, respectively; $V^*$ is only located in the $k$-subcore, where $k$ is the lower core number of two vertices that the inserted or removed edge connect.

\subsection{The Order-Based Core Maintenance}
\noindent
The state-of-the-art core maintenance solution is the \alg{Order} algorithm~\cite{Zhang2017,guo2022simplified}. For edge insertion, it is based on three notions: $k$-order, candidate degree, and remaining degree.
For edge removal, it uses the notion of the max-core degree \cite{sariyuce2013streaming}.

\subsubsection{Edge Insertion} 
\noindent
\begin{definition}[$k$-Order $\prec$]
{\cite{Zhang2017}}
Given a graph $G$, the $k$-order $\prec$ is defined for any pairs of vertices $u$ and $v$ over the graph $G$ as follows: (1) when $u.\core < v.\core$, we have $u \prec v$; (2) when $u.\core = v.\core$, we have $u \prec v$ if $u$'s core number is determined before $v$'s by the peeling steps of the BZ algorithm. 
\end{definition}

A $k$-order $\prec$ is an instance of all the possible vertex sequences produced by the BZ algorithm. When generating the $k$-order, there may be multiple vertices $v\in Q$ that have the same value of $u.d$ and can be dequeued from $Q$ at the same time together (Algorithm \ref{alg:bz}, line 4). When dealing with these vertices with the same value of $d$, different sequences generate different instances of correct $k$-order for all vertices. 
There are three heuristic strategies, ``small degree first'', ``large degree first'', and ``random''. The experiments in~\cite{Zhang2017} show that the ``small degree first'' consistently has the best performance over all tested real graphs, and thus we choose this strategy for implementation and experiments. 

For the $k$-order, transitivity holds, that is, $u\prec v $ if $u\prec w \land w\prec v$. For each edge insertion and removal, the $k$-order will be maintained.  
Here, $\od_k$ denotes the sequence of vertices in $k$-order whose core numbers are $k$. 
A sequence $\od = \od_0\od_1\od_2 \cdots$ over $V(G)$ can be obtained, where $\od_i \prec \od_j$ if $i < j$. It is clear that $\prec$ is defined over the sequence of $\od = \od_0\od_1\od_2\cdots$. In other words, for all vertices in the graph, the sequence $\od$ indicates the $k$-order $\prec$. 

Given an undirected graph $G=(V,E)$ with $\od$ in $k$-order, each edge $(u, v)\in E(G)$ can be assigned a direction such that $u \prec v$ and the directed edge is denoted as $u\mapsto v$.
By doing this, a \emph{direct acyclic graph} (DAG) $\vec G=(V,\vec E)$ can be constructed where each edge $u\mapsto v\in \vec E(\vec G)$ satisfies $u\prec v$. 
Of course, the $k$-order of $G$ is a {topological order} of $\vec G$.
Here, the set of successors of $v$ is defined as $u(\vec G).post=\{v \mid u\mapsto v \in \vec E (\vec G)\}$;
the set of predecessors of $v$ is defined as $u(\vec G).pre=\{v \mid v\mapsto u \in \vec E (\vec G)\}$.
When the context is clear, we use $u.\post$ and $u.\pre$ instead of $u(\vec G).\post$ and $u(\vec G).\pre$, respectively~\cite{guo2022simplified}.

\begin{definition} [candidate in-degree]
\rm{\cite{guo2022simplified,Zhang2017}}
Given a constructed DAG $\vec G(V,\vec E)$, the candidate in-degree $v.\degins$ is the total number of its predecessors located in $V^*$, denoted as $\degins(v) = |\{w\in v.pre: w\in V^*\}|$.
\end{definition}

\begin{definition} [remaining out-degree]
\rm{\cite{guo2022simplified,Zhang2017}}
Given a constructed DAG $\vec G(V,\vec E)$, the remaining out-degree $v.\degoutp$ is the total number of its successors without the ones that are confirmed not in $V^*$, denoted as $v.\degoutp = |\{w\in v.post: w\notin \vgr\}|$.
\end{definition}

\begin{theorem}
\label{th:potential-degree}
\rm 
{\cite{guo2022simplified}}
Given a constructed DAG $\vec G=(V,\vec E)$ by inserting an edge $u\mapsto v$ with $K = u.core \leq v.core$, the candidate set $V^*$ includes all possible vertices that satisfy: 1) their core numbers equal to $K$, and 2) their total numbers of candidate in-degree and remaining out-degree are greater than $K$, formally
$\forall w \in V: w \in V^* \equiv (w.\core = K~\land~w.\degins + w.\degoutp > K)$ 
\end{theorem}

For all vertices $v$ in $\vec G$, we must ensure that $v.\core \leq v.\degoutp$. When inserting an edge $v\mapsto u$, the out-degree $v.\degoutp$ increases by $1$. If $v.\core > v.\degoutp$, edge insertion maintenance is required after adding $v$ to $V^*$. Theorem \ref{th:potential-degree} shows what qualified vertices should be added to $V^*$. In this case, $V^*$ and $V^+$ are maintained, which are used to calculate $v.\degins$ and $v.\degoutp$ when traversing $v$.  

Algorithm \ref{alg:insert} is the \alg{Order} insertion algorithm~\cite{guo2022simplified}.
As the precondition, we assume that $v.\core$, $v.\degoutp$, and $v.\degins$ for all vertices $v$ are correctly initialized. 
After inserting an edge $u\mapsto v$, we add $u$ to $V^*$ when $u.\degoutp > K = u.\core$ (line 2). 
The key idea is to repeatedly add vertices $w$ to $V^*$ such that $w.\degins + w.\degoutp > K$.
Importantly, the priority queue $Q$ is used to traverse all affected vertices in $k$-order (lines 4 - 6).
When traversing the affected vertices $w$ in $\od$, the value $w.\degins+w.\degoutp$ is the upper bound as we traverse $\vec G$ in topological order. There are two cases.
First, if $w.\degins+w.\degoutp > K$, we execute the \texttt{Forward} procedure to add $w$ into $V^*$; 
also, for each $w'\in w.\post$ with $w'.\core = K$, we add $w'.\degins$ by one and then add to $Q$ for propagation (line 7).
Second, if $w.\degins+w.\degoutp \leq K \land w.\degins >0$, we identify that $w$ must not be added to $V^*$; 
procedure \texttt{Backward} propagates $w$ to remove potential vertices $v$ from $V^*$ since $v.\degoutp$ or $v.\degins$ is decreased (line 8). 
All other vertices $w$ in $\od$ not in the above two cases are skipped. 
Finally, $V^*$ includes all vertices whose core numbers should be added by 1 (line 9). 
Of course, the $k$-order is maintained for inserting other edges (line 10).

\iftrue
\begin{algorithm}[tb]
\SetAlgoNoEnd
\caption{{{Insert}Edge}($\vec G, \od, u\mapsto v$)}
\label{alg:insert}
\DontPrintSemicolon
\SetKwInOut{Input}{input}\SetKwInOut{Output}{output}
\SetKwFunction{Min}{min}
\SetKwFunction{Forward}{Forward}
\SetKwFunction{Backward}{Backward}
\SetKwFunction{Ending}{Ending}
\SetKwFunction{DoDeg}{DoDeg}
\SetKwData{True}{true}
\SetKwData{False}{false}
\SetKwData{Black}{black}
\SetKwData{Dark}{dark}
\SetKwData{Gray}{gray}
\SetKwData{White}{white}
\SetKwData{Red}{red}
\SetKw{Break}{break}
\SetKw{With}{:}
\SetKwFunction{}{}
\SetKw{Continue}{continue}


$V^*, V^+, K \gets \emptyset, \emptyset, u.\core$\;
insert $u\mapsto v$ into $\vec G$ with $u.\degoutp \gets u.\degoutp+1$\;
\lIf{$u.\degoutp \leq K$}{\textbf{return}}

$Q \gets $ a min-priority queue by $\od$; $Q.\enqueue (u)$\;
\While(){$Q \neq \emptyset$}{ 
    $w \gets Q.dequeue()$\;
    \lIf{$w.\degins + w.\degoutp > K$}{
        \Forward{$w, V^*, V^+$}
    }
    \lElseIf{$w.\degins > 0$} {
        \Backward{$w, V^*, V^+$}
    }
}

\lFor{$w\in V^*$}{$w.\core \gets K + 1$; $w.\degins \gets 0$}
{To maintain the $k$-order, remove each $w\in V^*$ from $\od_K$ and insert $w$ at the beginning of $\od_{K+1}$ in $k$-order.}\;

\end{algorithm} 
\fi

\begin{figure*}[!t]
\centering
\includegraphics[width=0.9\linewidth]{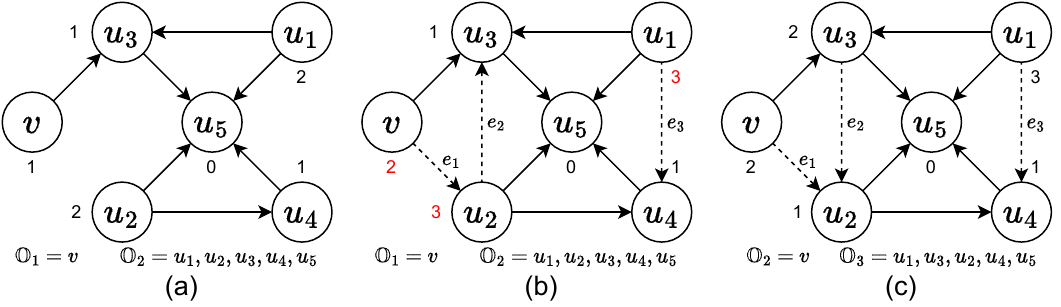} 
\centering
\caption{\rm An example graph maintains the core numbers after inserting three edges, $e_1$, $e_2$, and $e_3$ . The letters inside the cycles are vertices' IDs and the $\od_k$ is the $k$-order of vertices with core numbers $k$. The beside numbers are corresponding remaining out-degrees $\degoutp$. The direction for each edge indicates the $k$-order of two vertices, which is constructed as a DAG. 
(a) an initial example graph. (b) insert $3$ edges. (c) the core numbers and $k$-orders update. }
\label{fig:insert}
\end{figure*}

\begin{example}
\myfig\ref{fig:insert} shows an example of maintaining the core numbers of vertices after inserting three edges, $e_1$ to $e_3$, successively.
Figure \ref{fig:insert}(a) shows the example graph constructed as a DAG where the direction of edges indicates the $k$-order. After initialization, $v$ has a core number of $1$ with $k$-order $\od_1$ and $u_1$ to $u_5$ have a core number of $2$ with $k$-order $\od_2$. 

Figure \ref{fig:insert}(b) shows three edges, $e_1$, $e_2$ and $e_3$, being inserted. 
(1) For $e_1$, we increase $v.\degoutp$ to $2$ so that $v.\degoutp > v.\core$ and $V^*=\{v\}$. 
Then, we stop since all $v.\post$ have core numbers larger than $v.\core$. 
Finally, we increase $v.\core$ from $1$ to $2$.  
(2) For $e_2$, we increase $v.\degoutp$ to $3$ so that $v.\degoutp > v.\core$ and $V^*=\{u_2\}$. 
Then, we traverse $u_3$ in $k$-order and find that $u_3.\degins + u_3.\degoutp = 1 + 1 = 2 \leq K = 2$, so that $u_3$ cannot add to $V^*$ which cause $u_2$ removed from $V^*$ (by \texttt{Backward}). In this case, $u_2$ is moved after $u_1$ in $k$-order as $\od_2=u_1, u_3, u_2, u_4, u_5$. 
(3) For $e_3$, we increase $u_1.\degoutp$ to $3$ so that  $u_1.\degoutp > K = 2$ and $V^*=\{u_2\}$. Then, we traverse $u_3$, $u_2$, $u_4$ and $u_5$ in $k$-order, all of which can be added into $V^*$ since their $\degins+\degoutp > K = 2$. 
Finally, we increase the core numbers of $u_2$ to $u_5$ from $2$ to $3$. 

Figure \ref{fig:insert}(c) shows the result after inserting edges. All vertices have core numbers increased by $1$. Orders $\od_2$ and $\od_3$ are updated accordingly. All vertices' $\degoutp$ are updated accordingly. 
\end{example}

\subsubsection{Edge Removal}
\noindent
\begin{definition} [max-core degree $\mcd$] \cite{Saryuce2016,Zhang2017,guo2022simplified}
Given a graph $G=(V,E)$, for each vertex $v\in V$, the max-core degree is the number of $v$'s neighbors $w$ such that $w.core\geq v.core$, defined as $v.\mcd = |\{w\in v.\adj: w.\core \geq v.\core \}|$.
\end{definition}

All vertices $v$ in $G$ maintain $v.\mcd \geq v.\core$. When removing an edge $(u, v)$, e.g. $v.\core < u.\core$, we have $v.\mcd$ off by 1 and $u.\mcd$ unchanged. In this case, if $v.\mcd < v.\core$, Edge removal maintenance is required.  

The \alg{Order} removal algorithm is presented in Algorithm \ref{alg:orderremove}.
After an edge is removed, the affected vertices, $u$ and $v$, have to be put into $V^*$ if their $\mcd$ less than $\core$ (lines 2 to 4), which may repeatedly cause the other vertices' $\mcd$ to decrease and then be added to $V^*$ (lines 5 to 9). The queue $R$ is used to propagate the vertices added to $V^*$ whose $\mcd$ are less than their core numbers (lines 5 and 6). 
The $k$-order is maintained for inserting an edge next time (line 11).
Also, all vertices' $\mcd$ have to be updated for removing an edge next time (line 12). 

\begin{algorithm}[tb]
\SetAlgoNoEnd
\caption{{RemoveEdge}($G, \od, (u, v)$)}
\label{alg:orderremove}
\DontPrintSemicolon
\SetKwInOut{Input}{input}\SetKwInOut{Output}{output}
\SetKwFunction{Min}{Min}
\SetKwFunction{Forward}{Forward}
\SetKwFunction{Backward}{Backward}
\SetKwFunction{Ending}{Ending}
\SetKwFunction{DoDeg}{DoDeg}
\SetKwFunction{Delete}{{DELETE}}
\SetKwData{True}{true}
\SetKwData{False}{false}
\SetKwData{Black}{black}
\SetKwData{Dark}{dark}
\SetKwData{Gray}{gray}
\SetKwData{White}{white}
\SetKwData{Red}{red}
\SetKw{Break}{break}
\SetKw{With}{with}
\SetKwFunction{}{}
\SetKw{Continue}{continue}
    $R, K, V^* \gets$ an empty queue, $\Min(u.\core, v.\core), \emptyset$\;
    remove $(u, v)$ from $\vec G$ with updating $u.\mcd$ and $v.\mcd$\; 
    \lIf{$u.\mcd < K$}{$V^* \gets V^*\cup \{u\}; R.enqueue(u)$}
    \lIf{$v.\mcd < K$}{$V^* \gets V^*\cup \{v\}; R.enqueue(v)$}
    
    \While{$R\neq \emptyset$}{
        $w \gets R.dequeue()$\;
        
        \For{$w'\in w.\adj$ \With $w'.\core = K\land w \notin V^*$}{
            $w'.mcd \gets w'.mcd - 1$\;
            \lIf{$w'.\mcd < K$}{$V^* \gets V^*\cup \{w'\}; R.enqueue(w')$}
        }
    }
    
    \lFor{$w \in V^*$}{$w.\core \gets w.\core -1$}
    Remove all $w\in V^*$ from $\od_K$ and append to $\od_{K-1}$ in $k$-order\;
    update $\mcd$ for all related vertices accordingly\; 
\end{algorithm} 

\begin{figure*}[!t]
\centering
\includegraphics[width=0.9\linewidth]{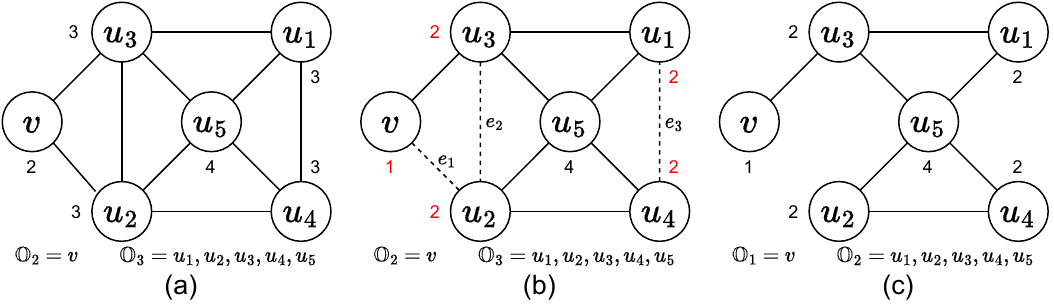} 
\centering
\caption{\rm An example graph maintains the core numbers after removing $3$ edges, $e_1$, $e_2$, and $e_3$. The letters inside the cycles are vertices' IDs and the $\od_k$ is the $k$-order of vertices with core numbers $k$. The beside numbers are corresponding $\mcd$. (a) an initial example graph. (b) remove three edges. (c) the core numbers and $\od_k$ update. }
\label{fig:remove}
\end{figure*}

\begin{example}
\myfig\ref{fig:remove} shows an example of maintaining the core numbers of vertices after successively removing three edges, $e_1$ to $e_3$.
Figure \ref{fig:remove}(a) shows that $v$ has a core number of 2 with $k$-order $\od_2$ and all $u_1$ to $u_5$ have core numbers of $3$ with $k$-order $\od_3$. For all vertices, the core numbers are less than or equal to $\mcd$.

\myfig\ref{fig:remove}(b) shows the three edges, $e_1$ to $e_3$, removed. (1) For $e_1$, $v.\mcd$ is off by $1$ so we have $v.\mcd < K = 2$ and $V^* = \{v\}$, but $u_2.\mcd$ is not affected. There is no propagation since all $v.\adj$ have core numbers greater than $K=2$. 
Finally, we decrease $v.\core$ from $2$ to $1$. 
(2) For $e_2$, both $u_2.\mcd$ and $u_3.\mcd$ are off by $1$ and less than $K=3$, so that $V^*=\{u_2, u_3\}$. Then, both $u_2$ and $u_3$ are added to $R$ for propagation, and $u_1$, $u_4$, and $u_5$ are consecutively added to $V^*$ with $V^*=\{u_2, u_3, u_1, u_4, u5\}$. 
Finally, we decrease the core number of $u_1$ to $u_5$ from $3$ to $2$; also, the $\mcd$ of both $u_2$ and $u_3$ are updated as $2$, and the $\mcd$ of $u_1$, $u_4$ and $u_5$ are updated as $3$.  
(3) For $e_3$, both $u_2.\mcd$ and $u_3.\mcd$ are off by $1$. But their $\mcd$ are still not less than $K=2$, so that $V^*=\emptyset$. The propagation stops. 
Finally, the $\mcd$ of both $u_1$ and $u_4$ are updated to $2$.

Figure \ref{fig:remove}(c) shows the result after removing the edges. All vertices have core numbers that decreased by $1$. Orders $\od_1$ and $\od_2$ are updated accordingly. 
Also, all vertices' $\mcd$ are updated accordingly. 
\end{example}

\subsection{Order Maintenance Data Structure}
\noindent
In the \alg{Simplified-Order} core maintenance algorithm~\cite{guo2022simplified}, the OM data structure~\cite{dietz1987two,bender2002two} is used to maintain the $k$-order. In this work, we adopt a concurrent version of the OM data structure~\cite{guo2022new} to maintain the $k$-order for the parallel core maintenance. The OM data structure has the following three operations: 
\begin{itemize}
    \item[--] $\texttt{Order}(\od, x, y)$: determine if $x$ precedes $y$ in the ordered list $\od$;
    \item[--] $\texttt{Insert}(\od, x, y)$: insert a new item $y$ after $x$ in the ordered list $\od$; 
    \item[--] $\texttt{Delete}(\od, x)$: delete $x$ from the total order in the ordered list $\od$.
\end{itemize}

Assume there are maximal $N$ items in the total order $\od$. All items are assigned labels to indicate the order.
For the \texttt{Insert} operation, a \emph{two-level} data structure~\cite{utterback2016provably} is used. Each item is stored in the bottom-list, which contains a \emph{group} of consecutive elements; each group is stored in the top-list, which can contain $\Omega(\log N)$ items. 
Both the top-list and bottom-list are organized as double-linked lists. We use $x.\pre$ and $x.\Next$ to denote the predecessor and successor of $x$, respectively.    
Each item $x$ has a top-label $L^t(x)$, which equals to $x$'s group label denoted as $L^t(x) = L(x.\group)$, and bottom-label $L_b(x)$, which is $x$'s label.   
With enough label space after $x$, $y$ can successfully obtain a new label in $O(1)$ time. Otherwise, the $x$'s group $g$ is \emph{full}, which triggers a \emph{relabel} process. 
\iftrue
Specifically, the relabel operations have two steps: 
\begin{itemize}
    \item[--] \emph{Rebalance}: if there is no label space after $x$'s group $g$, we have to rebalance the top-labels of groups. 
    From $g$, we continuously traverse the successors $g'$ until $L(g')-L(g)> j^2$, where $j$ is the number of traversed groups. Then, new group labels can be assigned with the gap $j$, in which newly created groups can be inserted. Finally, a new group can be inserted after $g$. 
    
    \item[--] \emph{Split}: when the group $g$ of $x$ is full, $g$ is split out one new group, which contains at most $\frac{\log N}{2}$ items and new bottom-labels $L_b$ are uniformly assigned for items in new groups. Newly created groups are inserted after $g$, where we can create the label space by the above rebalance operation. 
\end{itemize}
Such rebalance and split operations will continue until less than $\frac{\log N}{2}$ items remain in $g$. In addition, new bottom labels $L_b$ are uniformly assigned for items in $g$.  
\fi 

Label $L^t$ is in the range $[0, N^2]$ and label $L_b$ in the range $[0, N]$.
Typically, each label can be stored as an $O(\log N)$ bits integer. 
Assume it takes $O(1)$ time to compare two integers. For the sequential version~\cite{dietz1987two,bender2002two,utterback2016provably}, each \texttt{Insert} operation only cost amortized $O(1)$ time; also, the \texttt{Order} and \texttt{Delete} operations requires $O(1)$ time. 
In the parallel version~\cite{guo2022new}, \texttt{Insert} and \texttt{Delete} are synchronized by locks. More importantly, the parallel \texttt{Order} is lock-free, which is meaningful since a large portion of OM operations for core maintenance in graphs is to compare the order of two vertices on one edge. 

In this work, we adopt the parallel OM data structure~\cite{guo2022new} to maintain the $k$-order in parallel for three reasons. First, our method has a larger portion of \texttt{Order} operations compared to \texttt{Insert} and \texttt{Delete} operations. The lock-free \texttt{Order} operations are efficient even if multiple workers insert and remove vertices concurrently.
Second, all three operations cost $O(1)$ work, which does not worsen the work complexity of our core maintenance. 
Third, the labels of vertices, which indicate their order, can be used to implement the priority queue $Q$. Here, $Q$ is the key data structure for our core maintenance in Algorithm \ref{alg:p-edgeinsert}.

\subsection{Atomic Primitive and Lock}
\noindent
The compare-and-swap atomic primitive \texttt{CAS}$(x, a, b)$ takes a variable (location) $x$, an old value $a$ and a new value $b$. It checks the value of $x$, and if it equals $a$, it updates the variable to $b$ and returns \textit{true}; otherwise, it returns \textit{false} to indicate that updating failed. 

We use locks for synchronization in our parallel algorithms.  
In our experiments, we implemented two kinds of locks. One is the locks of OpenMP~\cite{mattson2019openmp}, \texttt{omp\_set\_lock} and \texttt{omp\_unset\_lock}, which is easy to use, but the overhead is high. In this paper, we use the compare-and-swap primitive (\texttt{CAS})~\cite{herlihy2020art}. 
We assume that the locks are weakly fair, and it guarantees that the threads are eventually executed if the condition is satisfied.

The \texttt{Lock} can be implemented by the atomic primitive \texttt{CAS}. 
Given variable $x$ as a lock, the \texttt{CAS} will repeatedly check $x$, and set $x$ from \texttt{false} to \texttt{true} if $x$ is \texttt{false}.
One worker will busy-wait on lock $x$ without suspension until another worker releases the lock.

Additionally, we use the conditional lock as in Algorithm \ref{alg:lockwith}. The {condition} $c$ is checked before and after the \texttt{CAS} primitive (lines 1 and 3). 
It is possible that other workers may update the condition $c$ concurrently. 
If $c$ is changed to \texttt{false} after locking $x$, variable $x$ will be unlocked and then return \texttt{false} immediately (line 4).
The conditional \texttt{Lock} can atomically lock $x$ by satisfying condition $c$, thus avoiding blocking on a locked $x$ that does not satisfy $c$. 
 
\begin{algorithm}[htb]
\SetKwData{True}{\small TRUE}
\SetKwData{False}{\small FALSE}
\SetKwData{True}{true}
\SetKwData{False}{false}
\SetKwFunction{CAS}{CAS}
\SetKw{Return}{return}
\caption{Lock($x$) with $c$}
\label{alg:lockwith}
\DontPrintSemicolon
\SetAlgoLined
\SetAlgoNoEnd
    \While{$c$}    {
        \If{$\CAS{x, \False, \True}$}{
        \lIf{$c$}{\Return \True}
        \lElse{$x \gets \False$; \Return \False}
        }
    }
    \Return \False
\end{algorithm}

\section{Parallel Core Maintenance}
\label{parallel-core-maint}
\noindent
The existing parallel core maintenance algorithms are based on the sequential \alg{Traversal} algorithm, which is proved to be less efficient than the sequential \alg{Order} algorithm. 
In this section, based on the \alg{Order} algorithm, we propose a new parallel core maintenance algorithm, \alg{Parallel-Order}, for both edge insertion and removal. 

The steps for parallel inserting edges are shown in Algorithm \ref{alg:parallel-insert}. 
Given an undirected graph $G$, the core number and $k$-order can be initialized by the \alg{BZ} algorithm~\cite{bz2003} in linear time. A batch of edges $\Delta E$ are to be inserted in $G$. We split these edges $\Delta E$ into $\wP$ parts, $\Delta E_1 \dots \Delta E_{\wP}$, where $\wP$ is the total number of workers (line 1). 
Each worker $p$ handles multiple edges in $\Delta E_p$ in parallel with other workers (line 2). 
Each time, a worker $p$ deals with a single edge in \texttt{InsertEdge}$_p$ (line 4).
The key issue is how to implement \texttt{InsertEdge}$_p$ executed by worker $p$ in parallel with other workers.

\iftrue 
\begin{algorithm}[htb]
\SetAlgoNoEnd
\caption{Parallel-InsertEdges($G, \od, \Delta E$)} 
\label{alg:parallel-insert}
\DontPrintSemicolon
\SetKwInOut{Input}{input}\SetKwInOut{Output}{output}
\SetKwFunction{DoInsert}{DoInsert$_1$}
\SetKwFunction{DoInsertP}{DoInsert$_{\mathcal{P}}$}
\SetKwFunction{DoInsertp}{DoInsert$_{p}$}
\SetKwFunction{EdgeInsert}{InsertEdge$_{p}$}
\SetKwData{True}{true}
\SetKwData{False}{false}
\SetKwData{Black}{black}
\SetKwData{Dark}{dark}
\SetKwData{Gray}{gray}
\SetKwData{White}{white}
\SetKwData{Red}{red}
\SetKw{Break}{break}
\SetKw{With}{$:$}
\SetKw{WITH}{with}
\SetKw{IN}{in}
\SetKwFunction{Lock}{Lock}
\SetKwFunction{Unlock}{Unlock}
\SetKw{Continue}{continue}
\SetKw{Break}{break}


partition $\Delta E$ into $\Delta E_1, \dots, \Delta E_{\wP}$\;
\DoInsert{$\Delta E_1$} $\parallel \dots \parallel$ \DoInsertP{$\Delta E_{\wP}$}

\medskip
\SetKwProg{myproc}{procedure}{ \DoInsertp{$\Delta E_p$}}{}
\myproc{}{
    \lFor{$(u, v) \in \Delta E_p$}{ \EdgeInsert{$G, \od, (u, v)$}
    }
}
\end{algorithm} 
\fi 

Removing edges in parallel is analogous to Algorithm~\ref{alg:parallel-insert}; the key issue is \texttt{RemoveEdge}$_p$. 
Note that the insertion and removal cannot run in parallel, which greatly simplifies the synchronization of the algorithms. 

One benefit of our method is that, unlike the existing parallel core maintenance methods~\cite{hua2019faster,Jin2018,wang2017parallel}, a prepossessing of $\Delta E_p$ is not required so that edges can be inserted and removed on-the-fly. 

\subsection{Parallel Edge Insertion}


\subsubsection{Algorithm}
\noindent
The detailed steps of \texttt{InsertEdge}$_p$ are shown in Algorithm~\ref{alg:p-edgeinsert}, which is analogous to Algorithm~\ref{alg:insert}. 
We introduce several new data structures. 
First, the priority queue $Q_p$, the queue $R_p$, the candidate set $V^*_p$, and the searching set $V^+_p$ are all privately used by the worker $p$; they cannot be accessed by other workers and synchronization is not necessary (lines 3, 7). 
Second, for each vertex $u\in V$, we introduce a status $u.s$, initialized to $0$ and atomically incremented by $1$ before and after the $k$-order operation (lines 16 and 30). 
In other words, when $u.s$ is an odd number, the $k$-order of $u$ is being maintained.
By using such a status for each vertex, we obtain $v \in u.\post$ ($u\prec v$) or $v \in u.\pre$ ($v\prec u$) by the parallel \texttt{Order}($u, v$) operation.

As shown in Algorithm \ref{alg:myorder}, when comparing the order of $u$ and $v$, we ensure that $u$ and $v$ are not updating their $k$-order. We repeatedly acquire $u.s$ and $v.s$ as $s$ and $s'$ until both $s$ and $s'$ are even numbers (line 3). After comparing the order of $u$ and $v$ (line 4), we check if $u.s$ and $v.s$ have increased (line 5). In that case, we redo the whole process (line 2). Finally, we return the result in line 6.  
Specifically, the status $s$ can be implemented by a 32-bit unsigned integer, which has two issues: First, a simple Boolean variable is not sufficient, as it must record the version number to ensure that there are no operations on $s$ during the comparison (line 4); second, an unsigned integer can overflow, e.g. the value $s = 2^{32} - 1$ adding by $1$ becomes $0$, but the comparison of $s$ (lines 3 and 5) is not affected, as a) the result of $(s~\texttt{mod}~2)$ is not changed even if $s$ overflows and b) it is impossible that $s$ increases $2^{32}$ times (during lines 2 - 5).

\iftrue
\begin{algorithm}[htb]
\small
\SetAlgoNoEnd
\caption{{{Parallel-Order}}($\od, u, v$)}
\label{alg:myorder}
\DontPrintSemicolon
\SetKwFunction{Even}{Even}
\SetKwFunction{Mod}{mod}
\SetKw{Goto}{goto}
\SetKwRepeat{Do}{do}{while}


$s\gets \varnothing; s'\gets \varnothing; r\gets \varnothing$\;
\Do(){$s \neq u.s~\lor~s' \neq v.s$}{
    \lDo(){$s~\Mod~2 = 1~\lor~s'~\Mod~2 = 1$}{
        $s \gets u.s$; $s' \gets v.s$
    }
    $r\gets u\prec v$\;
}

\Return $r$
\end{algorithm} 
\fi


Given an edge $u\mapsto v$ to be inserted, where $u\prec v$, we lock both $u$ and $v$ together at the same time when both are not locked (line~1);
if one is already locked, we must wait until both are unlocked. 
We redo the lock of $u$ and $v$ if they were updated by other workers as $v \prec u$ (line 2). 
After locking, $K$ is initialized as the smaller core number of $u$ and $v$.  
After inserting the edge $u \mapsto v$ in graph $G$ (line 4), $v$ can be unlocked (line 5). 
If $u.\degoutp \leq K$, we unlock $u$ and terminate (line 6); otherwise, we set $w$ to $u$ for propagation (line 7).
In the do-while-loop (lines 8 - 13), initially, $w$ equals $u$, which is already locked in line 1 (line 7).
We calculate $w.\degins$ by counting the number of $w.\pre$ located in $V^*_p$ while $w$ is locked since $w$ may be accessed by other workers (line 9). 
If $w.\degins + w.\degoutp > K $, vertex $w$ requires the \texttt{Forward} procedure$_p$ (line 10).
If $w.\degins + w.\degoutp \leq K \land w.\degins > 0$, vertex $w$ requires the \texttt{Backward} procedure (line 11). 
If $w.\degins + w.\degoutp \leq K \land w.\degins = 0$, we can skip $w$ and unlock $w$ since $w$ can not be in $V^+$ (line 11). 
Repeatedly, we dequeue $w$ from $Q_p$ (line 12) with the following steps: 1) we lock the head $w$ of $Q_p$; 
2) by checking $w.s$, we know whether $w$ has been locked and updated by other workers or not; 3) if that is the case, we remove $w$ from $Q_p$ if $w.\core > K$, unlock $w$ and update $Q_p$. 
In a word, we dequeue $w$ from $Q_p$, where $w$ has the minimal $k$-order in $Q_p$ and $w.\core = K$ (line 12). 
After locking $w$, its $k$-order cannot be changed by other workers. 
The do-while-loop terminates when no vertices can be dequeued from $Q_p$ (line 13).
All vertices $w \in V^*_p$ have core numbers incremented by $1$ and their $w.\degins$ is reset to $0$ (line 15); 
also, all vertices $w$ are removed from $\od_{K}$ and inserted at the head of $\od_{K+1}$ to maintain the $k$-order by using the parallel OM data structure, where $w.s$ is atomically incremented by $1$ before and after this process (line 16).
Before termination, all locked vertices $w$ are unlocked (line 17).  

Procedures \texttt{Forward}$(u)$ and \texttt{Backward}$(w)$ are almost the same as their sequential version since all vertices in $V^+$ are locked. There are only several differences.
In the \texttt{Forward}$_p(u)$ procedure, for each $v$ in $u.\post$ whose core number equals to $K$, we add $v$ in the priority queue $Q_p$ (line 21); but $v.\degins$ is not maintained by adding $1$ since it will be calculated in line 9 when using.  
In procedure \texttt{Backward}$_p(w)$, all vertices $w$ are removed from $\od_{K}$ and appended after $\pre$ to maintain the $k$-order by using the parallel OM data structure, where $w.s$ is atomically incremented by $1$ before and after this process (line 30).

\iftrue 
\begin{algorithm}[!htb]
\small
\SetAlgoNoEnd
\caption{InsertEdge$_p$($\vec G, \od, u\mapsto v$)}
\label{alg:p-edgeinsert}
\DontPrintSemicolon
\SetKwInOut{Input}{input}\SetKwInOut{Output}{output}
\SetKwFunction{Min}{min}
\SetKwFunction{Swap}{swap}
\SetKw{Goto}{goto}
\SetKwFunction{Forward}{Forward$_p$}
\SetKwFunction{Backward}{Backward$_p$}
\SetKwFunction{DoPre}{DoPre$_p$}
\SetKwFunction{DoPost}{DoPost$_p$}
\SetKwFunction{Insert}{Insert}
\SetKwFunction{Delete}{Delete}
\SetKwFunction{Ending}{Ending}
\SetKwFunction{DoDeg}{DoDeg}
\SetKwFunction{Lock}{Lock}
\SetKwFunction{Unlock}{Unlock}
\SetKwData{True}{true}
\SetKwData{False}{false}
\SetKwData{Black}{black}
\SetKwData{Dark}{dark}
\SetKwData{Gray}{gray}
\SetKwData{White}{white}
\SetKwData{Red}{red}
\SetKw{Break}{break}
\SetKw{With}{:}
\SetKw{Continue}{continue}
\SetKw{Return}{return}
\SetKwRepeat{Do}{do}{while}
\Lock $u$ and $v$ together if both are not locked\; 
\lIf{$v\prec u$}{\Unlock $u$ and $v$; \Goto line 1}
$V^*_p, V^+_p, K, \gets  \emptyset, \emptyset, \Min{u.\core, v.\core}$\;
insert $u\mapsto v$ into $\vec G$ with $u.\degoutp \gets u.\degoutp+1$\;
\Unlock{$v$}\;
\lIf{$u.\degoutp \leq K$}{
    \Unlock{u}; \Return
}

$Q_p, w \gets $ a min-priority queue by $\od, u$\;


\Do(){$w \neq \varnothing$}{ \label{alg:insert-while}
    $w.\degins \gets |\{w'\in w.\pre: w'\in V^*_p\}|$ \tcp*[r]{calculate $\degins$}
    \lIf{$w.\degins + w.\degoutp > K$}{
        \Forward{$w$}
    }
    \lElseIf{$w.\degins > 0$}{\Backward{$w$} \textbf{else} {\Unlock{$w$}} }
    
    \tcp{$w$ is locked and $w.\core = K$ when dequeuing $w$ from $Q_p$}
    $w\gets Q_p.\dequeue(K)$
}\label{alg:insert-while-end}

\For{$w\in V^*_p$}{
    $w.\core \gets K + 1$; $w.\degins \gets 0$\;
    \tcp{atomically add $w.s$}
    $\langle w.s\it{++}\rangle$; \Delete{$\od_K, w$}; \Insert{$\od_{K+1}, \it{head}, w$};
    $\langle w.s\it{++}\rangle$\;
}


\Unlock all locked vertices\;

\medskip
\SetKwProg{myproc}{procedure}{ \Forward{$u$}}{}
\myproc{}{
 $V^*_p \gets V^*_p \cup \{u\}$; $V^+_p\gets V^+_p \cup \{u\}$ \tcp*[r]{$u$ is locked}  

    \For{$v \in u.post: v.\core = K$}{
       
        \lIf{$v \notin Q_p$}{$Q_p.enqueue(v)$}
    }
}

\medskip
\SetKwProg{myproc}{procedure}{ \Backward{$w$}}{}
\myproc{}{
    $V^+_p \gets V^+_p \cup \{w\}$; $\pre \gets w$ \tcp*[r]{$w$ is locked} 
    $R_p \gets$ an empty queue;
    \DoPre{$w, R_p$}\;
    $w.\degoutp\gets w.\degoutp+ w.\degins$; $w.\degins \gets 0$\;    
    
    \While(){$R_p \neq \emptyset$}{ \label{backward:while}
    $u \gets R_p.\dequeue()$\;
    $V^*_p \gets V^*_p \setminus \{u\}$\; 
    \DoPre{$u, R_p$}; \DoPost{$u, R_p$}\;
    \tcp{atomically add $w.s$}
    $\langle w.s\it{++} \rangle$;
    \Delete($\od_K, u$); \Insert{$\od_K, \pre, u$}; 
    $\langle w.s\it{++} \rangle$\;  
    $\pre \gets u$;
    $u.\degoutp\gets u.\degoutp+ u.\degins$; $u.\degins \gets 0$\;
    } \label{backward:while-end}
}

\medskip
\SetKwProg{myproc}{procedure}{ \DoPre{$u, R_p$}}{}
\myproc{}{
    \For{$v \in u.\pre$ \With $v\in V^*_p$}{
        $v.\degoutp \gets v.\degoutp - 1$\;
        \lIf{$v.\degins + v.\degoutp \leq K~\land~v\notin R_p$}{
                $R_p.enqueue(v)$
        }
        
    }
}
\medskip
\SetKwProg{myproc}{procedure}{ \DoPost{$u, R_p$}}{}
\myproc{}{
    \For{$v \in u.\post$}{
            
            
            \If{$v \in V^*_p~\land~v.\degins > 0$}{
                $v.\degins \gets v.\degins -1$\;
                \lIf{$v.\degins + v.\degoutp \leq K~\land~v\notin R_p$} {
                     $R_p.enqueue(v)$
                }
            }
                
    }
}

\end{algorithm} 
\fi 


\begin{example}
Continuing in \myfig~\ref{fig:insert}, we show an example of maintaining the core numbers of vertices in parallel after inserting three edges. 
Figure \ref{fig:insert}(b) shows the insertion of three edges, $e_1$, $e_2$ and $e_3$ in parallel by three workers, $p_1$, $p_2$, and $p_3$, respectively. 
(1) For $e_1$, worker $p_1$ will first lock $v$ and $u_2$ for inserting the edge. But $u_2$ is already locked by $p_2$, so $p_1$ has to wait for $p_2$ to finish and unlock $u_2$. 
(2) For $e_2$, worker $p_2$ will first lock $u_2$ and $u_3$ for inserting the edge, after which $u_3$ is unlocked. Then, $u_3, u_4$ and $u_5$ are added to its priority queue $Q_2$ for propagation. That is, $u_3$ is locked and dequeued from $Q_2$ with $u_3.\degins = 1$ (assuming that $p_2$ locks $u_3$ before $p_3$ lock $u_3$). After propagation, we have that $V^*$ is empty. Continuing, $u_4$ and $u_5$ are locked and dequeued from $Q_2$, which are unlocked and skipped since their $\degins=0$. The $k$-order $\od_2$ is updated to $u_1, u_3, u_2, u_4$ and $u_5$.
(3) For $e_3$, worker $p_3$ first locks $u_1$ and $u_4$ for inserting the edge, after which $u_4$ is unlocked. 
Then, $u_3, u_4$ and $u_5$ are added to $Q_3$ for propagation. That is, $u_3$ is locked and dequeued from $Q_2$ (assuming that $p_3$ waits for $u_3$ unlocked by $p_2$) with $u_3.\degins = 1$, so $u_3$ is added to $V^*$ and $u_2$ is added to $Q_3$ for propagation. Continuing, $u_3$, $u_2$, $u_4$ and $u_5$ are locked and dequeued from $Q_3$ for propagation, which are all added to $V^*$ (assuming that $p_3$ waits for $u_2$ unlocked by $p_2$).

We can see how three vertices, $u_3$, $u_4$ and $u_5$, can be added to $Q_2$ and $Q_3$ at the same time. That is, when $p_3$ removes $u_3$ from $Q_2$, it is possible that $u_3$ has already been accessed by $p_2$. In this case, we have to update $Q_3$ before dequeuing if we find that $u_3$ is accessed by $p_2$; in case, the $k$-order of $u_3$ in $Q_3$ is changed by $p_2$.  

\end{example}

\subsubsection{Correctness}
We only argue the correctness of Algorithm \ref{alg:insert} related concurrency. There are no deadlocks since both $u$ and $v$ are locked together at the same time for an inserted edge $u\mapsto v$ (line 1), and the propagated vertices are locked in $k$-order (line 12). 

For each worker $p$, the accessed vertices are synchronized by locking. The key issue is to ensure that a vertex $w$ is locked and then dequeued from $Q_p$ in $k$-order in the do-while-loop (lines 8 - 12). The invariant is that $w$ has a minimal $k$-order in $Q_p$:
$$\forall v\in Q_p: w \notin Q_p \land w \prec v$$

Initially, the invariant is established as $w = u$ and $Q_p = \emptyset$. 
When dequeuing $w$ from $Q$, worker $p$ first locks $w$, which has the minimum in the $k$-order, and then removes $w$. In this case, other vertices $v\in Q$ can be accessed by other workers $q$. For this, there are two cases: 1) other vertices $v$ may have increased core numbers, which will be removed from $Q$; 2) other vertices $v$ may have $v.\degins + v.\degoutp \leq K$ and cannot be added to $V^*_{q}$, which may cause other vertices $v'$ to be removed from $V^*_q$ by procedure \texttt{Backward}; also, all vertices $v'$ are moved after $v$ in $k$-order, they cannot be moved before $v$. 
In other words, all vertices in $Q_p$ cannot have a smaller $k$-order than $w$ when $w$ is locked.

Worker $p$ traverse $u.\post$ in the for-loop (lines 20 - 21, 37 - 40), where $u$ is locked by $p$; but, not all $u.\post$ are locked by $p$ and may be locked by other workers for updating. So are $u.\pre$ in the for-loop (lines 33 - 35). 
The invariant is that all $u.\post$ have $k$-order greater than $u$ and all $u.\pre$ have $k$-order less than $u$: 
$$(v \in u.\post \implies u \prec v)~\land~(v'\in u.\pre \implies v' \prec u)$$
\begin{itemize}
    \item[--] $v \in u.\post \implies u \prec v$ is preserved as vertices $v$ may have increased core numbers, but $v$ will never be moved before $u$ in $k$-order by other workers $q$ by the \texttt{Backward} procedure, which has been proved before.
    
    \item[--] $v'\in u.\pre \implies v' \prec u$ is preserved as $u$ is already locked by worker $p$ so that no other workers can access $u$ and move $v'$ after $u$ in $k$-order by the \texttt{Backward} procedure. 
\end{itemize}
In other words, the set of $u.\post$ and $u.\pre$ will not change until $u$ is unlocked, even when other workers access the vertices in $u.\post$ and $u.\pre$.

\subsubsection{Time Complexity}
\noindent
When $m'$ edges are inserted into a graph, the total work is the same as that of the sequential version in Algorithm \ref{alg:insert}, which is $O(m'|E^+|\log |E^+|)$ where $E^+$ is the largest number of adjacent edges for all vertices in $V^+$ among each inserted edge, defined as $E^+ = \sum_{v\in V^+_p} v.\Deg$. 
In the best case, $m$ edges can be inserted in parallel by $\mathcal P$ workers with a depth $O(|E^+|\log |E^+| + m' |V^*|)$ as each worker will not be blocked by other workers; but all vertices in $V^*$ are removed from $\od_K$ and inserted sequentially at the head of $O_{K+1}$.   
Therefore, the best-case running time is $O(m' |E^+|\log |E^+|/\mathcal P + |E^+|\log |E^+| + m'|V^*|)$. 

In the worst case, $m'$ edges have to be inserted one by one, which is the same as total work, since $\mathcal P$ workers make a blocking chain. 
Therefore, the worst-case running time is $O(m' |E^+|\log |E^+|)$. 
However, in practice, such a worst-case does not always happen. The reason is that, given a large number of inserted edges, they have a low probability of connecting with the same vertex; also, each inserted edge has a small size of $V^+$ (e.g. 0 or 1) with a high probability, as shown in \myfig\ref{fig:vcolor}..

\subsubsection{Space Complexity}
For each vertex $v\in V$, it takes $O(3)$ space to store $v.\degins$, $v.\degoutp$, $v.s$, and locks, which take $O(3n)$ space in total.
Each worker $p$ maintains their private $V^*_p$, $V^+_p$, which takes $O(2|V^+| \mathcal P)$ space in total. 
Similarly, each worker $p$ maintains $Q_p$ and $R_p$, which take $O(|E^+| \mathcal P)$ space in total since at most $O(2|E^+|)$ vertices can be added to $Q_p$ and $R_p$ for each inserted edge. 
The OM data structure is used to maintain the $k$-order for all vertices in the graph, which takes $O(n)$ space. 
Therefore, the total space complexity is $O(n + |V^+| + |E^+| \mathcal P) = O(n + |E^+| \mathcal P)$.

\subsection{Parallel Edge Removal}



\subsubsection{Algorithm}
The steps of \texttt{RemoveEdge}$_p$ are shown in Algorithm \ref{alg:p-edgeremove}. 
We introduce several new data structures. 
First, the queue $R_p$ is privately used by worker $p$, which cannot be accessed by other workers and synchronization is not necessary (line 2).
Second, each worker $p$ adopts a set $A_p$ to record all the visited vertices $w' \in w.\adj$ to avoid repeatedly revisiting $w' \in w.\adj$ again.   
Third, each vertex $v\in V$ has a status $v.t$, which has four possible values:
\begin{itemize} 
    \item $v.t =  2$ means $v.\core$ is off by 1 add will be propagated by adding to $R_p$. Note that, the $v.\core = K -1$ and $v.t=2$ have to be executed atomically (line 22), as two values indicate one status. It can be simplify implemented by \texttt{CAS} lock, and the efficiency not likely to be affected since the operations in line 22 are short and quick. 
    \item $v.t =  1$ means $v$ is being propagated by the inner for-loop (lines 11 - 14). 
    \item $v.t = 3$ means  $v$ has to be checked again since some vertices $v.\adj$ have core numbers decreased by other workers.    
    \item $v.t = 0$ means $v$ is just initialized or already propagated. 
\end{itemize}

Given a removed edge $(u, v)$, we lock both $u$ and $v$ together at the same time when both are not locked (line 1); if one is already locked, we must wait until both are unlocked. 
After locking, $K$ is initialized to the smaller core numbers of $u$ and $v$ (line 2). 
We execute the procedure \texttt{CheckMCD}$_p$ for $u$ or $v$ to ensure that $u.\mcd$ and $v.\mcd$ are not empty (line 3). 
We remove the edge $(u, v)$ safely from the graph $G$ (line 4).
For $u$ or $v$, if their core number is greater than or equal to $K$, we execute procedure \texttt{DoMCD}$_p$ (lines 5 and 6), by which $u$ and $v$ may be added to $R_p$ for propagation.
If $u$ or $v$ are not in $R_p$, we immediately unlock $u$ or $v$ (line 7).
The while-loop (lines 8 - 16) propagates all vertices in $R_p$. 
A vertex $w$ is removed from $R_p$, and an empty set $A_p$ is initialized (line 9).
In the inner for-loop (lines 11 - 14), the adjacent vertices $w' \in w.\adj$ are conditionally locked with $w'.\core = K$ (line 11 and 12).  
Note that, to avoid deadlock, when locking $w'$, we will not busy-wait once the condition changed to $w'\core \neq K$, as $w'.\core$ can be decreased from $K$ to $K-1$ by other workers (line 12). 
For each locked $w'\in w.\adj$, we first execute procedure \texttt{CheckMCD}$_p$ in case $w'.\mcd$ is empty (line 15) and then execute the \texttt{DoMCD}$_p$ procedure (line 13); also, the visited $w'$ are added to $A_p$ to avoid visiting them repeatedly. 
We atomically decrease $w.t$ by $1$ before and after the inner for-loop since other workers can access $w.t$ in line 32 (lines 10 and 15).  
After that, if $w.t>0$, we have to propagate $w$ again as other vertices in $w.\adj$ have core numbers decreased from $K+1$ to $K$ by other workers (line 16).  
The while-loop will not terminate until $R_p$ becomes empty (line 8).
Finally, all vertices in $V^*$ are appended to $\od_{K-1}$ to maintain the $k$-order (line 17). 
All locked vertices are unlocked before termination (line 18). 

In procedure \texttt{DoMCD}$_p(u)$, vertex $u$ has been locked by worker $p$ (line 19). We decrease $u.\mcd$ by $1$, and $u.\mcd$ cannot be empty (line 20). 
If it still has $u.\mcd \geq u.\core$, we finally unlock $u$ and terminate (line 21 and 25).   
Otherwise, we first decrease $u.\core$ by $1$ and set $u.t$ as $2$ together, which has to be an atomic operation (line 22) since $v.t$ indicates $v$'s status for other workers. Here, line 22 can be implemented with \texttt{CAS} lock for simplicity.  
Then, we add $u$ to $R_p$ for propagation (line 23); also, we set $u.\mcd$ to empty since the value is out of date; it can be calculated later if used (line 24).

In procedure \texttt{CheckMCD}$(u)$, we recalculate $u.\mcd$ if it is empty (line 27). 
We initially set a temporary $\mcd$ to $0$ (line 28), and then count the $u.\mcd$ (lines 29 - 33).
Here, $u.\mcd$ is the number of $v\in u.\adj$ for two cases: 1) $v.\core \geq u.\core$, or 2) $v.\core = u.\core - 1$ and $v.t > 0$ (line 29); if that is the case, we increment the temporary $\mcd$ by one (line 30). 
When $v.\core = K - 1$, it is possible that $v.t$ is being updated by other workers. If $v.t$ equals $1$, we know that $v$ is being propagated. 
In this case, we have to set $v.s$ from $1$ to $3$ by the atomic primitive \texttt{CAS}, which leads to the propagation of $v$ to be redone in line 16 by other workers (line 32). 
We skip executing \texttt{CAS} when $v = w$ (line 32) to avoid many useless redo processes in line 13.   
If $v.t$ is reduced to $0$, the propagation of $v$ is finished so that $v$ cannot be counted as $u.\mcd$, and the temporary $\mcd$ is off by 1 (line 33).
Finally, we set $u.\mcd$ as the temporary $\mcd$ and terminate (line 34). 
The notable advantage is that we calculate $u.\mcd$ without locking $v \in u.\adj$.

\begin{algorithm}[htb]
\small
\caption{RemoveEdge$_p$($G,\od, (u,v)$)}
\label{alg:p-edgeremove}
\SetAlgoNoEnd
\DontPrintSemicolon
\SetKwInOut{Input}{input}\SetKwInOut{Output}{output}
\SetKwFunction{DELETE}{Parallel-DELETE}
\SetKwFunction{INSERT}{Parlalel-INSERT}
\SetKwFunction{Lock}{Lock}
\SetKwFunction{Unlock}{Unlock}
\SetKwFunction{Locktwo}{lock2}
\SetKwFunction{Unlocktwo}{unlock2}
\SetKwFunction{CheckMCD}{CheckMCD$_p$}
\SetKwFunction{PropagateMCD}{DoMCD$_p$}
\SetKwFunction{CAS}{CAS}
\SetKwFunction{Min}{Min}
\SetKwData{True}{true}
\SetKwData{False}{false}
\SetKwData{Black}{black}
\SetKwData{Dark}{dark}
\SetKwData{Gray}{gray}
\SetKwData{White}{white}
\SetKw{Continue}{continue}
\SetKw{Return}{return}
\SetKw{Goto}{goto}
\SetKw{With}{:}

 
\Lock $u$ and $v$ together if both are not locked\;
$K, R_p, V^*_p \gets$ \Min{$u.\core, v.\core$}, \text{an empty queue}, $\emptyset$\;
\CheckMCD{$u, \varnothing$}; \CheckMCD{$v, \varnothing$}\;
remove $(u, v)$ from $G$\; 
\lIf{$v.\core \geq K$}{ \PropagateMCD{$u$} }
\lIf{$u.\core \geq K$}{ \PropagateMCD{$v$} }

\Unlock $u$ if $u\notin R_p$; \Unlock $v$ if $v \notin R_p$\; 
\While(){ $R_p \neq \emptyset$ }{\label{alg:p-orderdelete-while}
    $w, A_p \gets R_p.\dequeue(), \emptyset$\;
    $\langle w.t\gets w.t - 1\rangle$  \tcp*{atomically sub} \label{alg:delete-redo}
    
    \For{$w' \in w.\adj:  w'\notin A_p \land w'.\core = K  $ }{
        \If{\Lock{$w'$} \texttt{with} $w'.\core = K$} {

         \CheckMCD{$w', w$}; \PropagateMCD{$w'$}\; $A_p\gets A_p \cup \{w'\}$ 
         }
        
    }
   
    $\langle w.t\gets w.t - 1\rangle$ \tcp*{atomically sub} 

    \lIf{$w.t > 0$}{\Goto line \ref{alg:delete-redo}}
    \label{alg:p-orderdelete-while-end}
}
Append all $u\in V^*_p$ at the tail of $\od_{K-1}$ in $k$-order\;
\Unlock all locked vertices \label{alg:p-edgeremove-unlock}

\medskip
\SetKwProg{myproc}{procedure}{ \PropagateMCD{$u$}}{}
\myproc{}{
   {$u.\mcd \gets u.\mcd -1$}\;
    \If{$u.\mcd < K $}{
        $\langle u.\core \gets K - 1$; 
        $u.t \gets 2 \rangle$ \label{alg:p-orderdelete-atomic} \tcp*{atomic operation}

        $R_p.\enqueue(u)$; $u.\mcd \gets \varnothing$\; 
        
        $V^*_p\gets V^*_p \cup \{u\}$; 
        \Delete{$\od, u$}
    }
    \lElse{\Unlock{$u$}}
}

\medskip
\SetKwProg{myproc}{procedure}{ \CheckMCD{$u, w$}}{}
\myproc{}{
    \lIf{$u.\mcd \neq \varnothing$}{\Return}
    $\mcd \gets 0$\;
    \For{$v\in u.\adj: v.\core \geq K \lor(v.\core = K-1 \land v.t > 0)$}{
        $\mcd \gets \mcd + 1$\;
        \If{$v.\core = K-1$}{
            \lIf{$v \neq w \land v.t = 1$}{\CAS{$v.t, 1, 3$}}
            \lIf{$v.t = 0$}{$\mcd \gets \mcd - 1$}
        }
    }
    $u.\mcd \gets \mcd$\;
    }

\end{algorithm}

\begin{example}
Continuing in \myfig\ref{fig:remove}, we show an example of maintaining the core numbers of vertices in parallel when removing three edges. 
Figure~\ref{fig:remove}(b) shows the removal of three edges, $e_1$, $e_2$ and $e_3$ in parallel by three workers, $p_1$, $p_2$ and $p_3$, respectively.
(1) For $e_1$, worker $p_1$ locsk $v$ and $u_2$ together for removing the edge. But $u_2$ is already locked by $p_2$, so $p_1$ has to wait for $p_2$ to unlock $u_2$. Then, $u_2$ is unlocked without changing $u_2.\mcd$; the core number of $v$ is off by $1$ and added to $R_1$ for propagation. Since only one vertex, $u_3 \in v.\adj$, has a core number greater than $v$, the propagation of $v$ terminates. Finally, $v$ is unlocked. 
(2) For $e_2$, worker $p_2$ first locks $u_2$ and $u_3$ together for removing the edge. 
Then, both $u_2.\core$ and $u_3.\core$ are off by $1$, and $u_2$ and $u_3$ are added to $R_2$ for propagation. 
For propagating $u_2$, we traverse all vertices $u_2.\adj$; vertex $u_4$ is locked by worker $p_3$. At the same time, $u_4.\core$ is decreased from $2$ to $1$, and $p_1$ skips to lock $u_4$ since the condition is not satisfied for the conditional lock.
Vertex $u_5$ is locked by $p_2$ and has $u_5.\mcd$ off by $1$. Similarly, for propagating $u_3$, we traverse all vertices $u_3.\adj$ by skipping $u_1$ and decreasing $u_5.\mcd$. Now, we have $u_5.\mcd = 2 < u_5.\core = 3$, so $u_5.\core$ is off by 1. Finally, we unlock $u_2, u_3$ and $u_5$; all their core numbers are $2$ now. 
(3) For $e_3$, worker $p_3$ first locks $u_1$ and $u_4$ together for removing the edge. Then both $u_1.\core$ and $u_4.\core$ are off by $1$, and $u_1$ and $u_4$ are added to $R_3$ for propagation. 
The propagation will stop since the neighbors of $u_1$ and $u_4$, $u_3, u_2$, and $u_5$ are locked by $p_2$ and have decreased core numbers.
Finally, we unlock $u_1$ and $u_4$; all their core numbers are $2$ now.
We can see that $p_2$ and $p_3$ execute without blocking each other, and all vertices in $V^*$ are locked.

The above example assumes that all vertices' $\mcd$ are initially generated. If $u_3.\mcd = \varnothing$ before removing $e_2$, we have to calculate $u_3.\mcd$ by \texttt{CheckMCD}. 
At this time, $u_2$ and $u_5$ are counted as $u_3.\mcd$ since they are not locked by $p_3$, but $u_1$ is locked by $p_3$ for propagation. The key issue is whether $u_1$ is counted as $u_3.\mcd$ or not. There are two cases: (1) if $u_1.\core = 3$, we increment $u_3.\mcd$ by $1$; (2) if $u_1.\core$ is decreased to $2$ and $u_1$ is propagating, we also increment $u_3.\mcd$ by $1$; since it is possible that $u_1$ has propagated $u_3$, we have to force $u_1$ to redo the propagation (setting $u_1.t$ from $1$ to $3$ atomically).

\end{example}

\subsubsection{Correctness}

Algorithm \ref{alg:p-edgeremove} has no deadlocks. First, both $u$ and $v$ are locked together at the same time for a removed edge $(u, v)$ (line~1). So, it is impossible for two workers to lock $u$ and $v$, respectively, which can lead to deadlocks. Second, all vertices $w\in R_p$ are locked by the worker $p$ and $w.\core = K - 1$; also, the worker $p$ will lock all $w' \in w.\adj$ with $w.\core = K$ for propagation (lines~11 and 12).
There are four cases:
\begin{itemize} 
    \item[--] if all vertices $w'$ are not locked, there is no deadlocks;
    
    \item[--] if $w'$ is locked by another worker $q$ but $w'.\core$ is not decreased, there is no deadlock as $w'$ has no propagation and worker $p$ will wait until $w'$ is unlocked by $q$;
    
    \item[--] if $w'$ is locked by another worker $q$ and $w'$ always has $w'.\core > K$, there is no deadlock as $w'$ is skipped for traversing.  
    
    \item[--] importantly, if $w'$ is locked by another worker $q$ and $w'.\core$ is decreased from $K$ to $K-1$, there is no deadlock as $w$ has propagation stopped on $w'$ for $w'.\core = K-1$ and $w'$ has propagation stopped on $w$ for $w.\core = K-1$. 
    We use ``\texttt{Lock with}'' to conditionally lock $w'$ with $w'.\core = K$, which ensures to stop busy-waiting when $w'.\core$ decreases from $K$ to $K-1$.  
\end{itemize}


The key issue of Algorithm \ref{alg:p-edgeremove} is to correctly maintain the $\mcd$ of all vertices in the graph: 
$
\forall v\in V: v.\mcd = |\{w\in v.\adj : w.\core \geq v.\core \}|
\label{eq:mcd}
$

With this definition of $\mcd$, all vertices $v$ in the graph satisfy $v.\mcd \geq v.\core$; 
when removing an edge, $v$ with $v.\mcd < v.\core$ are repeatedly added to $V^*_p$ and have their core numbers decreased by $1$ in order to make $v.\mcd \geq v.\core$. 
After deleting one edge, the vertices with decreased core numbers are added into $R_p$ for propagation. The key issue is to argue the correctness of the while-loop (lines 8 to 16) for propagation. 

We first define some useful notations. 
For all vertices $v\in V$, we use $v.\lock$, a boolean value, to denote that $v$ is locked and $\neg v.\lock$ to denote $v$ is unlocked.
We use $R$ to denote the union of all propagation queues, $R = \cup_{p=1}^{\mathcal P} R_p$. A vertex $v\in R$ indicates $v$ is in one of the $R_p$ for worker $p$.

The invariant of this while-loop is that all vertices $w\in V$ maintain a status $w.t$ indicating if $w$ in $R$ or not; 
for all vertices $w\in R_p$, which are locked and added to $V^*$, their core numbers are off by $1$ and $mcd$ set as to empty (waiting to be recalculated). For all vertices $w\in V$, if $w.\mcd$ is not empty, $w.\mcd$ is the number of neighbors $u$ that have core numbers that are 1) greater or equal $w.\core$, or 2) equal to $w.\core - 1$ with $u$ in $R$ waiting to be propagated:
\begin{equation*} \begin{split} \fs
&(\forall w \in V: (w.t > 0  \equiv w\in R) \land (w.t = 0 \equiv w\notin R)) \\
&\land~(\forall w \in R_p: w.\core = K-1 \land w.\mcd = \varnothing \land w\in V^*_p\\ 
&\qquad \land w.\lock\land w.t > 0) \\
&\land~(\forall u\in V: u.\mcd \neq \varnothing \implies u.\mcd = |\{v \in u.\adj:  \\ 
&\qquad  v.\core \geq  u.\core~\lor~ (v.\core = u.\core -1 \land v\in R) \}|)
\end{split} \end{equation*}

The invariant initially holds as the vertices $u$ and $v$ may be added to $R_p$ to remove the edge $(u, v)$, and $u$ and $v$ are locked if added to $R_p$. 
We now argue that the while-loop preserves the invariant: 
\begin{itemize}
\item[--] $\forall w \in V: (w.s > 0 \equiv w\in R) \land (w.s = 0 \equiv w\notin R)$ is preserved as $w.t$ is set to $2$ and $w$ is added to $R$ at the same time by the atomic operation in line 22; also, $w.s$ is off to $0$ when $w$ is removed from $R_p$ for proportion.  

\item[--] $\forall w \in R_p: w.\core = K-1 \land w.\mcd = \varnothing \land w\in V^*_p \land w.\lock\land w.s > 0$ is preserved as when adding $w$ to $R_p$, $w.\core$ is off by $1$, $w.\mcd$ is set to empty, $w.t$ is set to $2$, and $w$ is added to $V^*$; also, $w$ is locked before being added to $R$. 

\item[--] $\forall u\in V: u.\mcd \neq \varnothing \implies  u.\mcd = |\{v\in u.\adj: v.\core \geq u.\core \lor (v.\core = u.\core -1 \land v\in R) \}|$ is preserved as when $u.\mcd$ are calculated by procedure \texttt{CheckMCD}, $u$ may have neighbors $v\in u.\adj$ whose core numbers are off by $1$ and are added to $R$ by other workers, and the propagation has not yet happened. 
Note that, the atomic operation in line 22 ensures that $u$ has the core number off by 1 and added to $R$ at the same time. 
\end{itemize}

At the termination of the while-loop, the propagation queue $R$ is empty so that all vertices $w \in V$ have $w.\mcd$ correctly maintained.

We now argue the correctness of the inner for-loop (lines 11 - 14), which is important to parallelism. 
There are two more issues with the inner for-loop. 
One is that $w'.\core$ may be decreased from $K+1$ to $K$ concurrently by other workers after visiting $w'$ (line 11), which may lead to some $w'$ that have $w'.\core$ decreased to $K$ being skipped.
The other is that $v.\core$ and $v.s$ may be updated concurrently by other workers (line 29).

We first define useful notations as follows.
For the inner for-loop (lines 11 - 14), we denote the set of visited neighbors of $w$ as $w.V$, so that $w.V = \emptyset$ before the for-loop, $w.V\subseteq w.\adj$ when executing the for-loop, and $w.V = w.\adj$ after the for-loop; also, we denote the set of $A_p$ as $w.A_p$. 
Note that we redo the for-loop if $w.t>0$ by resetting the $w.V$ to empty (line 16).
We use $V^*$ to denote the union of all $V^*_p$, formally $V^* = \cup_{p=1}^{\mathcal P} V^*_p$. A vertex $v\in V^*$ indicates $v$ is in one of the $V^*_p$ for worker $p$. 

The invariant of the outer while-loop (lines 8 - 16) is preserved.
The additional invariant of the inner for-loop is that for all vertices $u\in V$, if $u.\mcd$ is not empty, $u.\mcd$ is the number of neighbors $v$ that have core numbers that are 1) greater or equal to $u.\core$, 2) equal to $u.\core - 1$ with $u\in R$, or 3) $w.\core - 1$ which has $u$ removed from $R$ for propagation but $v$ is not yet propagated by $u$. The status $v.t = 1$ indicates that $v$ is doing the propagation and $v.t = 0$ indicates that $v$ has finished the propagation:
\begin{equation*} \begin{split} \fs
&\forall w\in V: (w.t=1\lor w.t =3 \equiv w.V\subseteq w.\adj) \\
&\land~(\forall u\in V: u.\mcd \neq \varnothing \implies u.\mcd = |\{v \in u.\adj: \\ 
&\qquad v.\core \geq u.\core~\lor~(v.\core = u.\core -1 \land v\in R)~\lor \\
&\qquad (v.\core = u.\core -1\land v\notin R\land v\in V^* \land v\notin u.V\\
&\qquad \qquad  \land v \notin u.A_p)\}|)
\end{split} \end{equation*}

The invariant initially holds as we have $w.t=1 \land w.V=\emptyset \land w.A_p = \emptyset$. We now argue that the inner for-loop (lines 11 - 14) preserves the invariant:
\begin{itemize}
\item[--] $\forall w\in V: (w.t=1\lor w.t =3 \equiv w.V\subseteq w.\adj)$ is preserved as $w.t$ is set to $2$ when $w$ is added to $R_p$ and $w.s$ is off by 1 before and after the for-loop; also, $w.t$ may be atomically incremented by $2$ by \texttt{CAS} when a neighbor $w'$ in $w.\adj$ is calculating its $\mcd$. 

\item[--] $\forall u\in V: u.\mcd \neq \varnothing \implies u.\mcd = |\{v \in u.\adj:
v.\core \geq u.\core~\lor~(v.\core = u.\core -1 \land v\in R)~\lor 
(v.\core = u.\core -1 \land v\notin R \land v\in V^* \land v\notin u.V\land v \notin A_p) $ is preserved as when $u.\mcd$ is calculated by procedure \texttt{CheckMCD}, vertex $u$ may have neighbors $v\in u.\adj$ whose core numbers are off by 1 and added to $R$ for further propagation; also, it is possible that $v$ is added to $V^*$ and has already been removed from $R$ before the inner for-loop (line 9) for propagation; there are three cases:
\begin{enumerate}     
    \item if $v$ not yet traversed $u$ such that $v.\core = u.\core-1$ for propagation as $u \notin v.A_p$, $u$ should count $v$ as $u.\mcd$. 
    \item if $v$ has already traverse $u$ such that $v.\core = u.\core-1$ for propagation as $u\in v.A_p$, $u$ should not count $v$ as $u.\mcd$. 
    \item if $v$ has already traverse $u$ such that $v.\core \neq u.\core -1$ for propagation, but after that $u.\core$ has been updated to $v.\core = u.\core -1$, $u$ should count $v$ as $u.\mcd$. 
\end{enumerate}
The third case requires repeated traversing of $u$. 
We use $v.t = 1$ to let $u$ know that $v$ is executing the inner for-loop (lines 11 - 14) for propagating all vertices in $v.\adj$.  
When $v.t=1$, vertex $u$ will atomically increment $v.t$ by $1$ (line 32), and the propagation of $v$ can run again with $v.A_p$ avoiding repeated propagation (line 16). 
\end{itemize}


At the termination of the inner for-loop by $w.t = 0$, we have $w.V = w.\adj$, so the invariant of the while-loop holds. 
At the termination of the outer while-loop, the propagation queue $R$ is empty, so that all vertices $v\in V$ have $v.\mcd$ correctly maintained as in Equation \ref{eq:mcd}.

\subsubsection{Time Complexity}
When $m'$ edges are removed from the graph, the total work is the same as the sequential version in Algorithm \ref{alg:orderremove}, which is $O(m'|E^*|)$ where $E^*$ is the largest number of adjacent edges for all vertices in $V^*$ among each removed edge, defined as $E^* = \sum_{v\in V^*_p} v.\Deg$.
Similarly to edge insertion, in the best case, $m$ edges can be removed in parallel by $\mathcal P$ workers with a depth $O(|E^*|+m'|V^*|)$ as each worker will not be blocked by other workers; but all vertices in $V^*$ are removed from $\od_K$ and appended sequentially at the tail of $\od_{K-1}$. 
Therefore, the best-base running time is $O(m' |E^*|/\mathcal P +|E^*| + m'|V^*|)$.

Similarly to edge insertion, in the worst case, $m'$ edges have to be removed one by one, which is the same as total work, when $\mathcal P$ workers make a blocking chain. Therefore, the worst-case running time is $O(m'|E^*|)$.
However, in practice, such a worst-case does not always happen. The reason is that, given a large number of inserted edges, they have a low probability of connecting with the same vertex; also, each inserted edge has a small size of $V^*$ (e.g. 0 or 1) with a high probability, as shown in \myfig\ref{fig:vcolor}.


\subsubsection{Space Complexity}
For each vertex $v\in V$, it takes $O(1)$ space to store $v.\mcd$ and locks, so the space in total is $O(n)$.
Each worker $p$ maintains a private set $V^*_p$, which takes $O(|V^*| \mathcal P)$ space in total.
Each worker $p$ maintains a private set $R_p$, which takes $O(|E^*| \mathcal P)$ space in total since at most $O(|E^*|)$ vertices can be added to $R_p$ for each removed edge. 
The OM data structure is used to maintain the $k$-order for all vertices in the graph, which takes $O(n)$ space. 
Therefore, the total space complexity is $O(n + |V^*| \mathcal P + |E^*| \mathcal P) = O(n + |E^*| \mathcal P)$.

\section{Implementation}
\label{implementation}


\noindent
The min-priority queue $Q$ is used in core maintenance for edge insertion to efficiently obtain a vertex $v\in Q$ with a minimum $k$-order $\od$, where $\od$ is maintained by the parallel OM data structure.  
The queue $Q$ is implemented with a min-heap to compare the labels maintained by the parallel OM data structure, which supports the enqueuing and dequeuing in $O(\log |Q|)$ time.

The key issue is how to efficiently implement the enqueue and dequeue operations, when the vertices' labels in $\od_k$ are updated by the \texttt{relable} procedure (including rebalance and split) in the OM data structure.
The solution is to re-make the min-heap of $Q$ each time when the relable procedure is triggered in $\od_k$.
For the implementation, we require the following structures:
\begin{itemize} 
    \item[--] Each $k$-order $\od_k$ maintains a version number $\od_k.\version$, which is atomically incremented by $1$ before and after one triggered \texttt{relable} procedure. It can be implemented as a 32-bit unsigned integer and can overflow (like $2^{32} = 0$), but it is significantly large enough to support different versions.
    
    \item[--] Each $k$-order $\od_k$ maintains a counter $\od_k.\counter$, which can atomically record how many workers are executing the triggered relable procedure. 

    \item[--] The vertex $v$ is added to $Q$ along with the current top-label (group label), bottom-label, status and version number, denoted as $[L_b(v), L^t(v), v.s, \version]$. These two labels are used for min-priority in $Q$.

    \item[--] All vertices $v\in Q$ should have the same version number, which equals to $Q$'s version number $Q.\version$. 
\end{itemize}

\begin{definition} [Version Invariant]
All vertices $v$ in $Q$ maintain the invariant that all $v.\version$ are the same version numbers as $Q.\version$, denoted as $\forall v\in Q: v.\version = Q.\version$.
\label{def:version}
\end{definition}

The \texttt{dequeue} operation preserves the Version Invariant for all vertices $v$ in $Q$, as an inconsistent version number of vertices may lead to incorrect results. 
The steps of updating $Q.\version$ are shown in Algorithm \ref{alg:updateversion}. 
Initially, we set $\version'$ as the current version of $\od_K$ (line 1). 
If $ver'\neq Q.\version$, all the vertices $v$ in $Q$ will have their $[L_b(v), L^t(v), v.s, \version']$ updated to the current new values, with $\version'$ as their version numbers (lines 4 - 7).
We have to ensure that $\od_k.cnt = 0$ and $\version' = \od_k.\version$ during such a updating; otherwise, we will redo the update (lines 2 and 8). 
We also have to ensure that $v.s$ is an even number and is not changed during updating; otherwise, we have to redo the updating (lines 5 and 7) since other workers have accessed the vertices in $Q$ and their $k$-order may be changed. 
In other words, no other workers can execute during the updating process.  
Finally, we set $Q.\version$ to $\version'$ since all vertices in $Q$ have the same version as $\version'$ (line 6).


\begin{algorithm}[!htb]
\SetAlgoNoEnd
\SetKwData{True}{true}
\SetKwFunction{Await}{Await}
\SetKwFunction{Even}{Even}
\SetKwFunction{Odd}{Odd}
\SetKwFunction{Update}{update}
\SetKwData{False}{false}
\SetKw{Continue}{continue}
\SetKw{Break}{break}
\SetKw{Goto}{goto}
\SetKw{Return}{return}
\caption{$Q.\updateversion(\od_k)$}
\label{alg:updateversion}
\DontPrintSemicolon
\SetKwInOut{Input}{input}\SetKwInOut{Output}{output}

    $\version' \gets \od_K.\version$\; 
    \lIf{$\od_k.\counter \neq 0~\lor~\version' \neq \od_K.\version$}{ 
        \Goto line 1 
    }
    
    \If{$\version' \neq Q.\version$}{
        \For{$v \in Q$}{
        $s' \gets v.s$ \;
        Update $v$ with current $[L_b(v), L^t(v), s', \version']$\; 
        \lIf{$\neg$\Even{$s'$} $\lor~s' \neq v.s$}{ \Goto line 5}
        }
    }    
    
    \lIf{$\od_K.\counter \neq 0~\lor~\version' \neq \od_K.\version$}{\Goto line 1}
    $Q.\version \gets \version'$\; 
\end{algorithm} 

\subsubsection{Enqueue} 
The detailed steps of the \texttt{enqueue} operation are shown in Algorithm \ref{alg:enqueue}. 
We set $\version'$ to the current version of $\od_K$ (line 1). 
For the vertex $v$, we add the values of $[L_b(v), L^t(v), v.s, \version']$ to $Q$, with $\version'$ as their version numbers (line 2), and then update $Q$. 
If $\version'$ is not consistent with $\od_k.\version$ or $Q_p.\version$, we set $Q_p.\version$ to $\varnothing$, which indicates no version is assigned and the delayed version updating when executing \texttt{dequeue} operations.

\begin{algorithm}[!htb]
\small
\SetAlgoNoEnd
\SetKwData{True}{true}
\SetKwFunction{Await}{Await}
\SetKwFunction{Even}{Even}
\SetKwFunction{Odd}{Odd}
\SetKwData{False}{false}
\SetKw{Continue}{continue}
\SetKw{Break}{break}
\SetKw{Goto}{goto}
\SetKw{Return}{return}
\caption{$Q.\enqueue(\od_k, v)$}
\label{alg:enqueue}
\DontPrintSemicolon
\SetKwInOut{Input}{input}\SetKwInOut{Output}{output}

$\version', s' \gets \od_K.\version, v.s$\;
{Add $v$ into $Q$ with $[L_b(v), L^t(v), v.s, \version']$ and then update $Q$} \;
\If{$\version' \neq \od_K.\version \lor \version' \neq Q.\version \lor s' \neq s \lor \neg \Even{s}$}{
    $Q.\version \gets \varnothing$
}
\end{algorithm} 

\subsubsection{Dequeue}
Algorithm \ref{alg:dequeue} shows the steps of the dequeue operation. If $Q.\version$ is empty (not assigned in Algorithm~\ref{alg:enqueue} line 4), we update the version of $Q$ so that all vertices in $Q$ have consistent labels (line 2). 
In this case, we obtain $v$ as $Q.\it{front}()$. which has the lowest $k$-order by comparing the labels (line 3). 
We conditionally lock $v$ with $v.\core = k$ as we will skip $v$ when it has $v.\core \neq k$ for the core maintenance (lines 4 and 5), since $v$ can be accessed by other workers and has an increased core number. 
After locking $v$, it is necessary to check $v$'s current status value $v.s$ with $v$'s status value in $Q$ (lines 6 and 7).
If they are not equal, we know that $v$ has been accessed by other workers, $v.\core = k$, and $v$'s $k$-order may be changed; then $Q.\version$ is set to empty to update the version in the next round (line 7).
We remove $v$ from $Q$ and then return $v$, which is locked with the smallest $k$-order with $v.\core = k$ (line 11).
The whole process continues until we successfully obtain $v$ from $Q$ or $Q$ is empty (lines 1 and 8). 
If no qualified $v$ exists in $Q$, it will return empty (line 9).

\begin{algorithm}[!htb]
\SetAlgoNoEnd
\SetKwData{True}{true}
\SetKwFunction{Await}{Await}
\SetKwFunction{Even}{Even}
\SetKwFunction{Odd}{Odd}
\SetKwData{False}{false}
\SetKw{Continue}{continue}
\SetKw{Break}{break}
\SetKw{Goto}{goto}
\SetKw{Return}{return}
\caption{$Q.\dequeue(\od_k)$}
\label{alg:dequeue}
\DontPrintSemicolon
\SetKwInOut{Input}{input}\SetKwInOut{Output}{output}

\While{$Q \neq \emptyset$}{
    \lIf{$Q.\version = \varnothing$}{$Q_p.\updateversion(\od_k)$}
    $v \gets Q.\it{front}()$\; 
    \If{$\neg (\Lock~v~\texttt{with}~v.\core = k)$} {Remove $v$ from $Q$; \Continue}

    \If{$v.s \neq [v.s]_{Q}$}{ 
        \Unlock{$v$}; $Q_p.\version \gets \varnothing$; \Continue
    } 
    Remove $v$ from $Q_p$; \Return $v$\; \label{alg:dequeue-return}
    
}
\Return $\varnothing$\;



\end{algorithm} 

\subsubsection{Running time}
The priority queue can be implemented by min-heap, which requires worst-case $O(\log |Q|)$ time for both enqueuing and dequeuing one item. 
For our implementation, with version updating, the priority queue requires worst-case $O(\log|Q|)$ for enqueuing and $O(|Q| \log |Q|)$ for dequeuing,  as we may rebuild the min-heap each time when removing a vertex. 
However, such a worst-case can happen with a low probability for three reasons. 
First, practically, the vertices are always inserted into the different positions of $\od_K$ with a high probability.
Second, only a limited number of relable procedures can be triggered for the Order Data Structure, since the amortized running time for the insert operation is bounded by $O(1)$. 
Third, for inserted edges, practically, the sizes of $Q$ are always small, e.g., 0 and 1, as shown in \myfig\ref{fig:vcolor}; so the process of updating version tends to not affect the dequeue performance.

\section{Experiments}
\label{experiments}
\noindent
In this section, we experimentally compare our parallel core maintenance approach with the state-of-the-art \emph{join-edge-set} core maintenance approach~\cite{hua2019faster}. 
There are four algorithms: 
\begin{itemize} 
    \item[--] our parallel edge insertion algorithm (\texttt{OurI} for short) and removal algorithm (\texttt{OurR} for short),
    \item[--] the \emph{join edge set} based parallel edge insertion algorithm (\texttt{JEI} for short) and removal algorithm (\texttt{JER} for short)
\end{itemize}

\subsection{Experiment Setup}
\noindent
The experiments are performed on a server with an AMD CPU (64 cores, 128 hyperthreads, 256 MB of last-level shared cache) and 256 GB of main memory. The server runs the Ubuntu Linux (22.04) operating system. All tested algorithms are implemented in
C++ and compiled with g++ version 11.2.0 with the -O3 option. OpenMP \footnote{\url{https://www.openmp.org/}} version 4.5 is used as the threading library. 
Our \texttt{OurI} and \texttt{OurR} use locks for synchronization, which are implemented by the \texttt{CAS} primitive for busy waiting.  
We perform every experiment at least 50 times and calculate their means with 95\% confidence intervals.

\subsection{Tested Graphs}
\noindent
We evaluate the performance of different methods over a variety of real-world and synthetic graphs, which are shown in Table \ref{tb:graph}. 
For simplicity, directed graphs are converted to undirected ones in our testing; all of the self-loops and repeated edges are removed. That is, a vertex can not connect to itself, and each pair of vertices can connect with at most one edge. 
The \emph{livej}, \emph{patent}, \emph{wiki-talk}, and \emph{roadNet-CA} graphs are obtained from SNAP\footnote{\url{http://snap.stanford.edu/data/index.html}}. 
The \emph{dbpedia}, \emph{baidu}, \emph{pokec} and \emph{wiki-talk-en} \emph{wiki-links-en} graphs are collected from the KONECT\footnote{\url{http://konect.cc/networks/}} project. 
The \emph{ER}, \emph{BA}, and \emph{RMAT} graphs are synthetic graphs; they are generated by the SNAP\footnote{\url{http://snap.stanford.edu/snappy/doc/reference/generators.html}} system using Erd\"{o}s-R\'enyi, Barabasi-Albert, and the R-MAT graph models, respectively. For these generated graphs, the average degree is fixed to 8 by choosing 1,000,000 vertices and 8,000,000 edges. 
All the above twelve graphs are static graphs, and we randomly sample edges for insertion and removal.

We also select four real temporal graphs, \emph{DBLP}, \emph{Flickr}, \emph{StackOverflow}, and \emph{wiki-edits-sh} from KONECT. For a temporal graph, each edge has a timestamp recording the time of this edge inserted into the graph. We select a batch of edges within a continuous time range for insertion and removal. 

\begin{table}[tb]
\centering
\caption{Tested real and synthetic graphs.}
\begin{tabular}{l|rrrr}
\toprule
Graph & $n=|V|$ & $m=|E|$ & AvgDeg & Max $k$ \\ 
\midrule
livej & 4,847,571 & 68,993,773 & 14.23 & 372 \\
patent & 6,009,555 & 16,518,948 & 2.75 & 64 \\
wikitalk & 2,394,385 & 5,021,410 & 2.10 & 131 \\
roadNet-CA & 1,971,281 & 5,533,214 & 2.81 & 3 \\ \midrule
dbpedia & 3,966,925 & 13,820,853 & 3.48 & 20 \\
baidu & 2,141,301 & 17,794,839 & 8.31 & 78 \\
pokec & 1,632,804 & 30,622,564 & 18.75 & 47 \\
wiki-talk-en & 2,987,536 & 24,981,163 & 8.36 & 210 \\
wiki-links-en & 5,710,993 & 130,160,392 & 22.79 & 821 \\ \midrule
ER & 1,000,000 & 8,000,000 & 8.00 & 11 \\
BA & 1,000,000 & 8,000,000 & 8.00 & 8 \\
RMAT & 1,000,000 & 8,000,000 & 8.00 & 237 \\ \midrule
DBLP & 1,824,701 & 29,487,744 & 16.17 & 286\\
Flickr & 2,302,926 & 33,140,017 & 14.41 & 600 \\
StackOverflow & 2,601,977 & 63,497,050 &24.41& 198\\
wiki-edits-sh & 4,589,850 & 40,578,944 &8.84& 47 \\
\bottomrule
\end{tabular}

\label{tb:graph}
\end{table}

In Table \ref{tb:graph}, we can see that all graphs have millions of edges, their average degrees ranges from 2.1 to 22.8, and their maximal core numbers ranges from 3 to 821. 
For most graphs, the core numbers are not well distributed. That is, a great portion of vertices have small core numbers, and few have large core numbers. 
For example, \emph{wikitalk} has $1.7$ million vertices with a core number of $1$; \emph{roadNet-CA} has four core numbers from $0$ to $3$;  \emph{ER} has nine core numbers from $2$ to $11$; \emph{BA} only has a single core number of $8$.
For \texttt{JEI} and \texttt{JER}, the core number distribution of graphs is an important property since the vertices with the same core number can only be handled by one worker at the same time, e.g., in \emph{BA} only one worker can execute and all the other workers are wasted. However, \texttt{OurI} and \texttt{OurR} do not have such a limitation for parallelism, so all workers can always run in parallel over all tested graphs.

\subsection{Running Time Evaluation}

\begin{figure*}[!t]
\centering
\includegraphics[width=\linewidth]{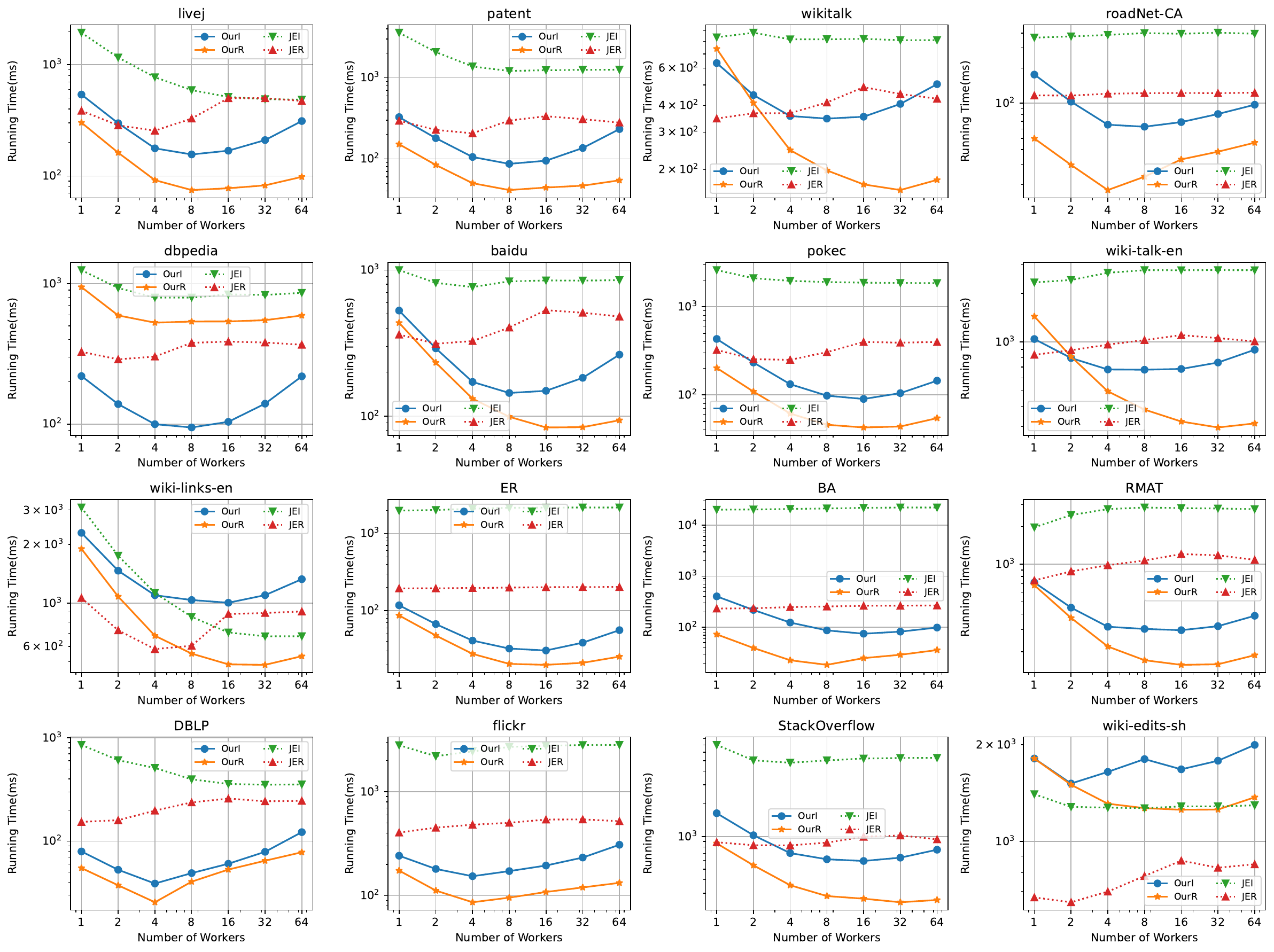} 
\centering
\caption{The real running time by varying the number of workers. The x-axis is the number of workers, and the y-axis is the execution time (millisecond). The error bars are too small to show.}
\label{fig:time}
\end{figure*}

\noindent
In this experiment, we exponentially increase the number of workers from $1$ to $64$ to evaluate the real running time over graphs in Table \ref{tb:graph}.
For each graph, we first randomly select 100,000 edges. 
We measure the accumulated running time for inserting or removing such 100,000 edges. 
The plots In \myfig\ref{fig:time} depict the performance of the four compared algorithms.
The first look over all tested graphs reveals that \texttt{OurI} and \texttt{OurR} always have better performance than \texttt{JEI} and \texttt{JER}, respectively. 
Specifically, we make several observations: 
\begin{itemize}
    \item[--] By using one worker, all algorithms are reduced to running sequentially, and  \texttt{OurI} performs much faster than \texttt{JEI}. 
    This is because for edge insertion, \texttt{OurI} is based on the \alg{Order} algorithm, while \texttt{JEI} is based on the \alg{Traversal} algorithm. It is proved that the \alg{Order} is much faster than \alg{Traversal} for sequential version \cite{Zhang2017}. Also, \texttt{JEI} requires the preprocessing time to generate the join edge sets, while \texttt{OurI} can run on-the-fly without preprocessing. 

    \item[--] By using one worker, \texttt{OurR} does not always perform better than \texttt{JER}. 
    This is because our method uses arrays to store edges, which can save space, while the join-edge-set-based method uses binary search trees to store edges. When deleting an edge $(u, v)$, \texttt{OurR} has to traverse all vertices of $u.\adj$ and $v.\adj$, while \texttt{JER} only needs to traverse $\log |u.\adj|$ and $\log |v.\adj|$ vertices. That means \texttt{OurR} has a higher running time than \texttt{JER} to delete an edge from the graph.  

    \item[--] By using multiple workers, \texttt{OurI} and \texttt{OurR} can always achieve better speedups compare with \texttt{JEI} and \texttt{JER}, respectively. 
    This is because \texttt{JEI} and \texttt{JER} have limited parallelism, as affected vertices with different core numbers can not be processed in parallel, while \texttt{OurI} and \texttt{OurR} do not have such a limitation.  
    On many real graphs, e.g., \emph{wikitalk}, \emph{roadNet-CA}, \emph{ER}, \emph{BA} and \emph{RMAT}, the core numbers are not well-distributed, and a large percent of vertices have the same core numbers. 
    Over such graphs, by increasing the number of workers, \texttt{JEI} and \texttt{JER} have no speedups since one worker needs to traverse a large subgraph, which takes a great amount of time,
    but \texttt{OurI} and \texttt{OurR} can always obtain speedups since all workers have almost equal probability to traverse the graphs. 
    
    \item[--] The running time of all tested algorithms may begin to increase when using more than 8 or 16 workers in certain graphs, e.g., \emph{livej},  \emph{patent}, and \emph{dbpedia}.
    This is because of the contention on the shared data structures with multiple workers, and more workers may lead to higher contention. 
    In addition, for \texttt{JEI} and \texttt{JER}, when the core numbers of vertices in graphs are not well distributed, some workers are wasted, which results in extra overheads.
\end{itemize}

\begin{table*}[!htb]
\centering
\caption{Compare the speedups.}
\begin{tabular}{l|cccc | cc | cc }
\toprule
   & \multicolumn{4}{c|}{1-worker vs 16-worker} & \multicolumn{2}{c|}{1-worker} & \multicolumn{2}{c}{16-worker} \\
Graph & \texttt{OurI} & \texttt{OurR} & \texttt{JEI}  & \texttt{JER} & 
    \texttt{OurI} vs \texttt{JEI}& \texttt{OurR} vs \texttt{JER} 
    & \texttt{OurI} vs \texttt{JEI} & \texttt{OurR} vs \texttt{JER} \\  
\midrule
livej & 3.2 & 3.9 & 3.8 & 0.8 & 3.6 & 1.3 & 3.0 & 6.5 \\
patent & 3.4 & 3.4 & 2.9 & 0.9 & 11.0 & 1.9 & 13.0 & 7.6 \\
wikitalk & 1.8 & 4.4 & 1.0 & 0.7 & 1.3 & 0.5 & 2.3 & 2.9 \\
roadNet-CA & 2.6 & 1.5 & 0.9 & 1.0 & 2.1 & 2.3 & 5.7 & 3.7 \\
dbpedia & 2.1 & 1.8 & 1.5 & 0.8 & 5.7 & \textbf{0.3} & 8.1 & \textbf{0.7} \\
baidu & 3.5 & \textbf{5.2} & 1.2 & 0.7 & 1.9 & 0.8 & 5.7 & 6.3 \\
pokec & 4.8 & 4.7 & 1.4 & 0.8 & 6.0 & 1.6 & 20.9 & 9.3 \\
wiki-talk-en & 1.5 & 4.5 & 0.8 & 0.8 & 2.2 & 0.6 & 4.1 & 3.4 \\
wiki-links-en & 2.3 & 3.9 & \textbf{4.4} & \textbf{1.2} & 1.3 & 0.6 & \textbf{0.7} & 1.8 \\
ER & 3.8 & 4.4 & 0.9 & 1.0 & 16.8 & 2.2 & 70.9 & 10.1 \\
BA & \textbf{5.4} & 2.9 & 0.9 & 0.9 & \textbf{49.6} & \textbf{3.2} & \textbf{289.1} & \textbf{10.6} \\
RMAT & 2.4 & 4.3 & \textbf{0.7} & \textbf{0.6} & 2.8 & 1.1 & 9.5 & 7.7 \\
DBLP & 1.3 & \textbf{1.0} & 2.4 & 0.6 & 10.8 & 2.8 & 5.9 & 4.9 \\
flickr & 1.2 & 1.6 & 1.0 & 0.7 & 11.6 & 2.3 & 14.3 & 5.0 \\
StackOverflow & 2.8 & 3.2 & 1.3 & 0.9 & 4.3 & 1.0 & 8.8 & 3.7 \\
wiki-edits-sh & \textbf{1.1} & 1.4 & 1.1 & 0.8 & \textbf{0.8} & 0.4 & 0.8 & 0.7 \\
\bottomrule
\end{tabular}

\label{tb:speedup}
\end{table*}

In Table~\ref{tb:speedup}, columns 2 to 5 compare the running time speedups between using one worker and 16 workers for all tested algorithms. It is clear that \texttt{OurI} and \texttt{OurR} consistently achieve better speedups up to 5x, compared with \texttt{JEI} and \texttt{JER}.    
Columns 6 to 9 compare the running time speedups between our method and the compared method using one worker or 16 workers. 
We can see that compared with \texttt{JEI}, \texttt{OurI} achieves up to a 50x speedup even using one worker, and achieves up to a 289x speedup when using 16 workers.
We observe that compared with \texttt{JER}, \texttt{OurR} does not always achieve speedups when using a single worker, but achieves up to 10x speedups when using 16 workers.
Especially, over \emph{wiki-edits-sh}, \texttt{OurI} and \texttt{OurR} run slower than \texttt{JEI} and \texttt{JER} when using 1 worker and 16 workers, respectively. The reason is that the special properties of graphs may affect the performance of our algorithms.


\subsection{Scalability Evaluation}

\begin{figure}[htb]
\centering
\includegraphics[width=\linewidth]{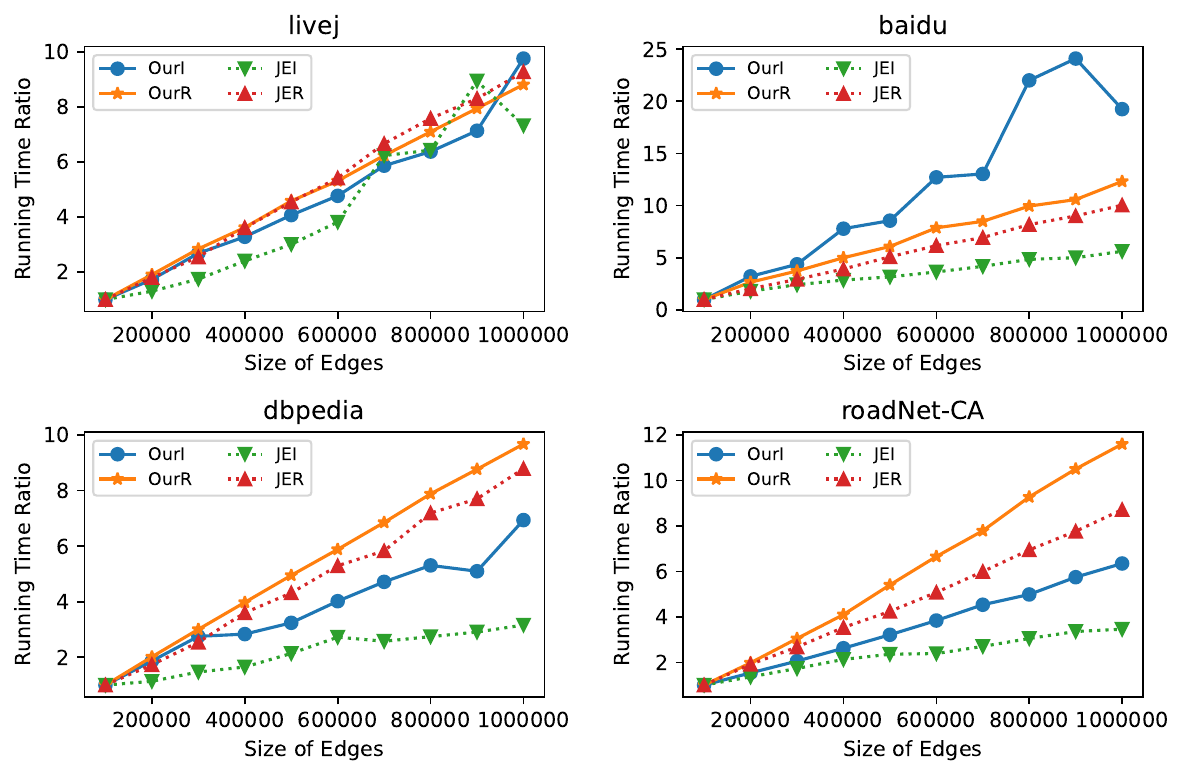} 
\centering
\caption{The running time ratio with 16 workers by varying the size of the inserted or removed edges. The x-axis is the size of inserted or removed edges, and the y-axis is the time ratio.}
\label{fig:time-size}
\end{figure}

\noindent
In this experiment, we test the scalability over four selected graphs, \emph{livej}, \emph{baidu}, \emph{dbpedia}, \emph{roadNet-CA}.
For each graph, we first randomly select 100,000 to 1 million edges.  
By using 16 workers, we measure the accumulated running time and evaluate the ratio of real running time between the corresponding size of edges and 100,000 edges.
The plots in \myfig\ref{fig:time-size} depict the performance of the four compared algorithms. Ideally, 1 million edges should have a ratio of 10 since the edge size is 10 times of 100,000.
We observe that over \emph{livej}, the four algorithms always have similar time ratios with increased an edge size. 
Over other graphs, \texttt{OurI} and \texttt{OurR} always have larger time ratios compared to \texttt{JEI} and \texttt{JER}, respectively. 
Further, \texttt{OurI} has a time ratio of up to 20 when applying 1 million edges. 
This is because \texttt{JEI} or \texttt{JER} adopts the joint edge set structure to preprocess a batch of updated edges; if there are more updated edges, they can process more edges in each iteration and avoid unnecessary access.
However, \texttt{OurI} and \texttt{OurR} do not preprocess a batch of updated edges, so more edges require more accumulated running time.  

We also observe that even with 1 million edges, \texttt{OurI} and \texttt{OurR} still have better performance than \texttt{JEI} and \texttt{JER}, respectively. 
Over four tested graphs, \texttt{OurI} still has 2.6x, 1.9x, 3.8x and 3.0x speedups compared with \texttt{JEI}, and 
\texttt{OuR} also has 7.8x, 5.5x, 0.9x and 3.3x speedups compared to \texttt{JER}, respectively. 
The reason is that \texttt{OurI} and \texttt{OurR} (based on the \alg{Order} algorithm) have less work than \texttt{JEI} and \texttt{JER} (based on the \alg{Traversal} algorithm; also, unlike \texttt{OurI} and \texttt{OurR}, \texttt{JEI} and \texttt{JER} have extra cost to preprocess the edges.

\subsection{Stability Evaluation}

In this experiment, we test the stability over four selected graphs, \emph{livej}, \emph{baidu}, \emph{dbpedia}, \emph{roadNet-CA}, by using 16 workers.
First, we randomly sample 5, 000, 000 edges and partition them into 50 groups, where each group has totally different 100, 000 edges. Second, for each group, we measure the accumulated running time of different methods.
That is, the experiments run 50 times, and each time has totally
different inserted or removed edges.

\begin{figure}[htb]
\centering
\includegraphics[width=\linewidth]{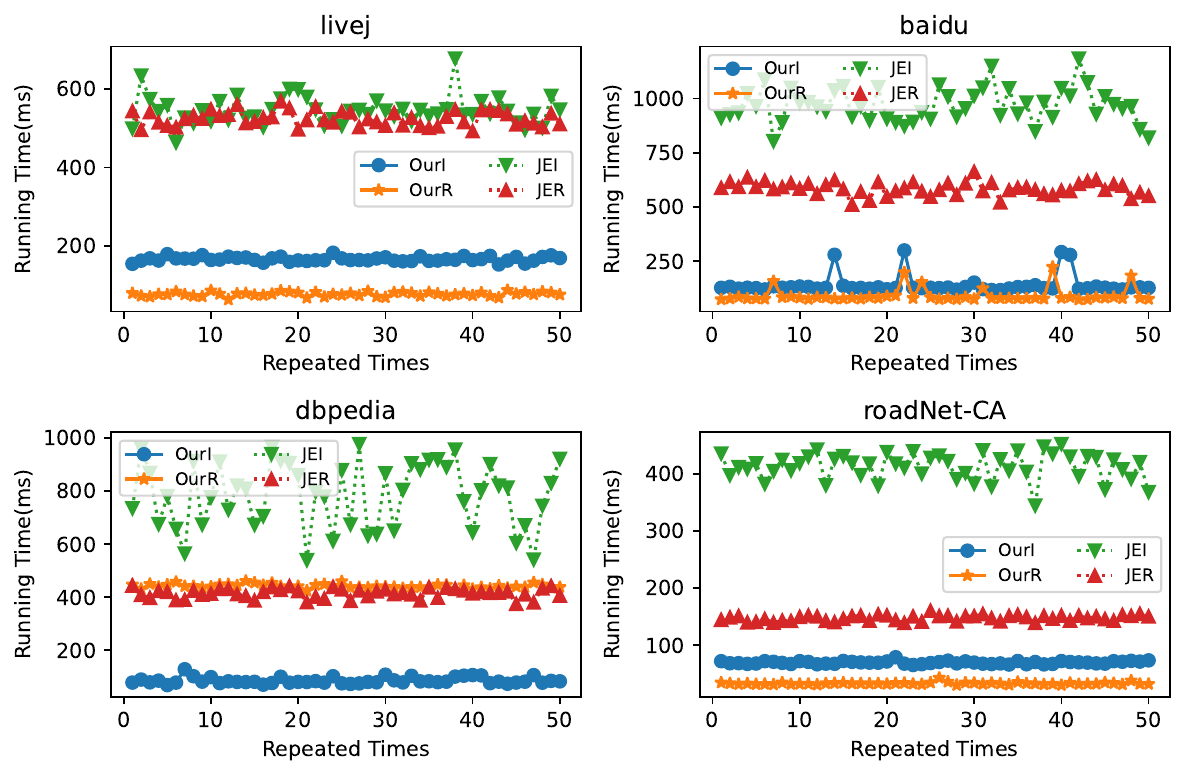} 
\centering
\caption{\rm The running time with 16-worker by varying a batch of inserted or removed edges for each time. The x-axis is the repeated times, and the y-axis is the running times.}
\label{fig:time-stability}
\end{figure}

The plots in \myfig\ref{fig:time-stability} depict the result. 
We observe that the performance of \texttt{OurI}, \texttt{OurR}, and \texttt{JER} are always well-bounded, but the performance of \texttt{JEI} has larger fluctuations. The reason is that \texttt{JEI} is based on the \alg{Traversal} algorithm and \texttt{OurI} is based on the \alg{Order} algorithm. 
It is proved that for edge insertion, the \alg{Traversal} algorithm has a large fluctuations ratio of $|V^+|/|V^*|$ for different edges with high probability, while the \alg{Order} algorithm does not have this problem. 
For edge removal, both \texttt{OurR} and \texttt{JER} have $V^+ = V^*$, so their performance remains stable for different batches of edges.

\section{Conclusions and Future Work}
\label{conclusions}
\noindent
We present new parallel core maintenance algorithms for inserting and removing a batch of edges based on the \alg{Order} algorithm. A set $V^+$ of vertices is traversed. We use locks for synchronization. Only the vertices in $V^+$ are locked, and their associated edges are not necessarily locked, which allows high parallelism.

The proposed parallel methodology can be applied to other graphs, e.g.~weighted graphs and probability graphs. 
It can also be applied to other graph algorithms, e.g.~maintaining the $k$-truss in dynamic graphs. Additionally, the maintenance of the hierarchical $k$-core involves maintaining the connections among different $k$-cores in the hierarchy, which can benefit from our result.  

\bibliographystyle{elsarticle-num}
\bibliography{references.bib}
\end{document}